\documentclass[12pt,twoside,onecolumn,draftcls]{IEEEtran} 
\usepackage{graphicx,cite,amsmath,amssymb,hhline}
\usepackage[utf8]{inputenc}
\usepackage{verbatim}

\begin{document}

\title{Outage Probability of Dual-Hop Selective AF With Randomly Distributed and Fixed Interferers}


\author{
\bigskip
\medskip {\normalsize $\mbox{Kezhi Wang}^{}$, {\em Student Member, IEEE,} $\mbox{Yunfei Chen}^{}$, {\em Senior Member, IEEE,}
    $\mbox{Marco Di Renzo}^{}$,  {\em Member, IEEE}}
\thanks{Kezhi Wang and Yunfei Chen
    are with the School of Engineering,
    University of Warwick, Coventry, U.K. CV4 7AL
    (e-mails: Kezhi.Wang@warwick.ac.uk, Yunfei.Chen@warwick.ac.uk)}
\thanks{M. Di Renzo is with the Laboratoire des Signaux et Systèmes, Unité
Mixte de Recherche 8506, Centre National de la Recherche Scientifique-École
Supérieure d’Électricité-Université Paris-Sud XI, 91192 Gif-sur-Yvette Cedex,
France (e-mail: marco.direnzo@lss.supelec.fr).}\\
    }
\thispagestyle{empty} \setcounter{page}{0}
\maketitle

\renewcommand{\baselinestretch}{1.73} \normalsize

\newpage
\begin{abstract}
The outage probability performance of a dual-hop amplify-and-forward
selective relaying system with global relay selection is analyzed
for Nakagami-$m$ fading
channels in the presence of multiple interferers at both the relays
and the destination. Two different cases are considered. In the
first case, the interferers are assumed to have random number and locations. Outage probability using the generalized Gamma
approximation (GGA) in the form of one-dimensional integral is
derived. In the second case, the interferers are
assumed to have fixed number and locations. Exact outage probability in the form of one-dimensional integral is derived. For both cases, closed-form expressions of lower bounds and asymptotic expressions for high signal-to-interference-plus-noise ratio are also provided.
Simplified closed-form expressions of outage probability for special cases (e.g., dominant interferences, i.i.d. interferers, Rayleigh distributed signals) are studied. Numerical results are
presented to show the accuracy of our analysis by examining the effects of the number and locations of interferers on the outage performances of both AF systems with random and fixed interferers.
\end{abstract}



\section{Introduction}
\label{one}
Wireless relaying can extend the network coverage by using idle nodes as relays in the network. It can also provide diversity gain by using idle nodes as "virtual" antennas \cite{904590}. Consequently, a huge amount of works have been conducted on its application in future wireless networks. Among all the relaying strategies, amplify-and-forward (AF) and decode-and-forward (DF) are perhaps the most widely used ones. In AF relaying, the source broadcasts its information to the relays in the first phase and then the relays simply amplify the received signals from the source and forward them to the destination in the second phase, while in DF relaying, the source broadcasts its information in the first phase but the relays have to decode the received signals from the source and then re-encode the signals before forwarding them to the destination. Due to its lower complexity, AF relaying is more attractive for some applications \cite{1374901}. On the other hand, in practical systems, it is often the case that more than one idle nodes are available at the same time such that multiple relays can be used by the source. If all the idle nodes are used in wireless relaying, orthogonal channels between relays have to be used in the second relaying phase such that the relayed signals will not interfere with each other. This will reduce the usage efficiency of the system resources considerably. To solve this problem, relay selection can be used by choosing only one node out of all available idle nodes in the second phase \cite{5876342}. It can be shown that relay selection can achieve the same diversity gain as the scheme that uses all available idle nodes, with proper designs \cite{4801494}. Thus, this paper focuses on relay selection using AF as an effective technology to achieve reliable communications.

Several researchers have studied the performance of relay selection
using AF. In \cite{4290052}, the optimal relay selection criterion was
proposed by selecting the relay with the largest instantaneous
end-to-end or global signal-to-noise ratio (SNR) for forwarding. The
performance of this criterion was analyzed in \cite{4020532}. In
\cite{1603719}, two suboptimal relay selection schemes based on two
upper bounds to the instantaneous global SNR were proposed and
analyzed. Reference \cite{4489652} proposed partial relay selection
scheme where relay selection is based on only the instantaneous SNR
of the first hop. In \cite{5876342}, both the optimal selection scheme
and the partial selection scheme were analyzed for Nakagami-$m$
fading channels. Two new partial relay selection schemes were also
proposed in \cite{5876342}. However, none of these works considered the interferences from other
transmitting sources in the network. In
a multiple-access system or a frequency-reused cell, interferences
from other transmitting sources, such as interferers, may cause performance degradation
and therefore, cannot be ignored. Moreover, the positions of the nodes may not be optimized such that interferers may be randomly distributed. In this case, the spatially averaged (over the distributions of the positions) performance metrics may be of more practical use for system design and optimization by considering random locations of interferers.
Reference \cite{6555163} provided the closed-form expression of the outage probability of dual–hop AF relaying in the presence of interference
at the destination over Rayleigh fading channels. Reference \cite{6130607} analyzed the outage probability of a dual-hop AF relaying system where both relay and
destination are interfered by a single source in Nakagami-$m$ fading channel.
However
these two works either considered interferers only at the destination over Rayleigh fading channels or only a signal interferer of fixed location.

In this paper, we provide a comprehensive analytical framework to
derive the outage probability performance of an AF relay selection
system where the relays suffer from path loss, independent but
non-identically distributed Nakagami-$m$ fading as well as Nakagami-$m$ interferences. In the first case, the
interferers have random number and locations. This is the case for multiple-access systems with mobile nodes. In the second
case, the interferers have fixed number and locations. This
is the case for fixed-access wireless systems where wireless
interconnections are mainly provided to replace wires with
considerably low or little mobility. The optimal criterion that selects the relay
with the largest instantaneous global signal-to-noise-plus-interference ratio (SINR) is studied. The exact outage
probability in terms of either one-dimensional integrals or closed-form approximations are derived. Also, lower bounds to outage probability are given. Finally,
asymptotic expression of outage probability in Rayleigh fading are studied for large SINR values.
Numerical results are presented to show the accuracy of our analysis
and therefore to examine the effects of interferences
on relay selection using AF.

The remainder of this paper is organized as follows. Section
\uppercase\expandafter{\romannumeral2} introduces the system model.
Section \uppercase\expandafter{\romannumeral3} considers the case when the interferers have random number and
locations, while Section
\uppercase\expandafter{\romannumeral4} studies the case when the interferers have fixed number and
locations. Numerical results are presented in Section
\uppercase\expandafter{\romannumeral5}, followed by concluding
remarks in Section \uppercase\expandafter{\romannumeral6}.
\section{System Model}
\label{two} Consider a wireless relaying system with one source $S$,
one destination $D$ and multiple relays $J$ of $R_j$, $j=1,2,\cdots,J$.
There is no direct link
between the source and the destination, which is the case when relays are used to extend network coverage and is the focus of this paper. All nodes have a single antenna and are in half-duplex mode.
Assume that there are $I^{sj}$
interferers of $I^{sj}_i$, $i=1,2,\cdots,I^{sj}$, and $I^{jd}$
interferers of $I^{jd}_v$, $v=1,2,\cdots,I^{jd}$,
that are transmitting at the
same time as the source to the $j$-th relay and the $j$-th relay to the destination, causing interferences to the
$j$-th relays and the destination, respectively. Assume that the distance between the source $S$ and the $j$-th relay
$R_j$, the $j$-th relay $R_j$ and the
destination $D$, the $i$-th interferer $I^{sj}_i$
and the $j$-th relay $R_j$ and the $v$-th
interferer $I^{jd}_v$ and the destination $D$ are $l_{sj}$, $l_{jd}$,
$l_{ij}$ and $l_{vj}$, respectively. Also, assume that the path loss between $S$ and
$R_j$, $R_j$
and $D$,
$I^{sj}_i$ and $R_j$, $I^{jd}_v$ and $D$ are $\eta(l_{sj})$,
$\eta(l_{jd})$, $\eta(l_{ij})$ and $\eta(l_{vj})$, respectively. As
the singular path loss model leads to impractical power condition in
the network when $l<1$, we assume the non-singular model for the
path loss as
\begin{equation}\label{ri1}
\begin{aligned}
\eta(l)=\frac{1}{l^\beta +1}
\end{aligned}
\end{equation}
where $\beta$ is the path loss exponent.

In the case when the interferers have random number and locations, we assume the numbers $I^{sj}$, $I^{jd}$ and the distances $l_{ij}$, $l_{vj}$ are random. We assume a Poisson point process
(PPP) with density $\lambda_I$ for the spatial distribution of the
interferers. Then, the probability density function (PDF) of the
number of interferers is given as
\begin{equation}\label{ri2}
\begin{aligned}
Pr\left\{i=I\right\}=\frac{(\lambda_I A_I)^I}{I!}e^{-\lambda_I
A_I},\;I=0,1,\cdots
\end{aligned}
\end{equation}
where $A_I$ is the distribution area of interferers. Also, we assume that the distance $l$ follows a general distribution with a PDF of
$f_l(x)$ which can be specified for different applications
considered.
In the case
when the interferers have fixed locations, both the number of
interferers and their locations are fixed such that $I^{sj}$, $I^{jd}$, $l_{ij}$ and $l_{vj}$ are deterministic values.

The received signal from the source $S$ to the $j$-th relay $R_j$ is given by
\begin{equation}
\label{ri3}
y_{sj} = \sqrt{\Omega_{sj}}h_{sj}x + \sum_{i=1}^{I^{sj}}\sqrt{\Omega_{ij}}h_{ij}x_{ij} + n_{sj}
\end{equation}
and it can be further amplified and
forwarded such that the received signal at the destination is given by
\begin{equation}
\label{ri4} y_{jd} = \sqrt{\Omega_{jd}}h_{jd}\cdot G_j\cdot
y_{sj} + \sum_{v=1}^{I^{jd}}\sqrt{\Omega_{vj}}h_{vj}x_{vj} + n_{jd}
\end{equation}
where $\Omega_{sj}=K_{sj}P_{sj}|h_{sj}|^2\eta(l_{sj})$, $\Omega_{jd}=K_{jd}P_{jd}|h_{jd}|^2\eta(l_{jd})$, $\Omega_{ij}=K_{ij}P_{ij}|h_{ij}|^2\eta(l_{ij})$ and $\Omega_{vj}=K_{vj}P_{vj}|h_{vj}|^2\eta(l_{vj})$
are the average power of
the Nakagami-$m$ fading gain in the channel between the source $S$ and the $j$-th relay
$R_j$, the $j$-th relay $R_j$ and the
destination $D$, the $i$-th interferer $I^{sj}_i$
and the $j$-th relay $R_j$, the $v$-th
interferer $I^{jd}_v$ and the destination $D$, respectively, $P_{sj}$, $P_{jd}$, $P_{ij}$ and $P_{vj}$
are the transmitted power of $S$, $R_j$, $I^{sj}_i$ and $I^{jd}_v$, respectively, $K_{sj}$, $K_{jd}$, $K_{ij}$ and $K_{vj}$
are constants that take other power factors, such as antenna gains and the average signal
power factors, into account, $h_{sj}$, $h_{jd}$, $h_{ij}$ and $h_{vj}$ are the fading
gains with unit average power between $S$ and
$R_j$, $R_j$
and $D$,
$I^{sj}_i$ and $R_j$, $I^{jd}_v$ and $D$, respectively, $x$, $x_{ij}$ and $x_{vj}$ are the transmitted symbol of $S$, $I^{sj}_i$ and $I^{jd}_v$, respectively,
$n_{sj}$, $n_{jd}$ are the additive white Gaussian noise (AWGN)
in the channel between $S$ and $R_j$, $R_j$
and $D$, respectively, and $G_j$ is the amplification
factor.

In the above, assume enough distances
between relays and between source and relays and between relays and destination
such that $|h_{sj}|$, $j=1,2,\cdots,J$, are independent of each other, $|h_{jd}|$, $j=1,2,\cdots,J$ are independent of each other and $|h_{sj}|$ are independent of $|h_{jd}|$. Similarly, we assume that $n_{sj}$ are independent of each other for different $j$ and $n_{jd}$ are independent of each other for different $j$ and $n_{sj}$ are independent of $n_{jd}$.
Also, assume
enough distances between interferers at the relay and between interferers at the destination such that
$I^{sj}_i$, $|h_{ij}|$, $i=1,2,\cdots,I^{sj}$, at $R_j$ are
independent of each other, respectively, and $I^{jd}_v$, $|h_{vj}|$, $v=1,2,\cdots,I^{jd}$, at $D$ are
independent of each other, respectively. We also assume interferers change from broadcasting phase to
relaying phase such that interferences at the destination are
independent of those at the relays. Note also that $I^{sj}_i$, $|h_{ij}|$ are independent of $I^{jd}_v$, $|h_{vj}|$, respectively, for different values of $j$, as it is not possible to have the same interferences in the signals from different relays to destination. Otherwise, the relays have to transmit their signals at the same time in the same frequency band and the destination will not be able to tell which signal is from which relay.

Based on discussions above, we assume independent Nakagami-$m$ fading channels such that the fading powers
$|h_{sj}|^2$, $|h_{jd}|^2$, $|h_{ij}|^2$, and $|h_{vj}|^2$ are
independent Gamma random variables with shape parameters $m_{sj}$,
$m_{jd}$, $m_{ij}$, $m_{vj}$ and scale parameters $1/m_{sj}$,
$1/m_{jd}$, $1/m_{ij}$, $1/m_{vj}$, respectively, where the Nakagami
$m$ parameters are assumed to be integers. Also, assume
$E\{|x|^2\}=1$, $E\{|x_{ij}|^2\}=1$ and $E\{|x_{vj}|^2\}=1$ such
that the actual average signal power is absorbed by $\Omega_{sj}$,
$\Omega_{ij}$ and $\Omega_{vj}$, respectively. Denote
$\sigma_{sj}^2=E\{|n_{sj}|^2\}$ as the noise power between $S$ and $R_j$ and
$\sigma_{jd}^2=E\{|n_{jd}|^2\}$ as the noise power between $R_j$ and $D$. Let
$Y_{sj}=\sum_{i=1}^{I^{sj}}\Omega_{ij}|h_{ij}|^2$ and
$Y_{jd}=\sum_{v=1}^{I^{jd}}\Omega_{vj}|h_{vj}|^2$.
Define the SNR between $S$ and $R_j$, $R_j$
and $D$ as $\gamma^{SNR}_{sj}=\frac{E\{\Omega_{sj}|h_{sj}|^2\}}{\sigma_{sj}^2}$,
$\gamma^{SNR}_{jd}=\frac{E\{\Omega_{jd}|h_{jd}|^2\}}{\sigma_{jd}^2}$, respectively, the interference-to-noise ratio (INR) between $S$ and $R_j$, $R_j$
and $D$ as $\gamma^{INR}_{sj}=\frac{E\{Y_{sj}\}}{\sigma_{sj}^2}$, $\gamma^{INR}_{jd}=\frac{E\{Y_{jd}\}}{\sigma_{jd}^2}$, respectively, and the SINR between $S$ and $R_j$, $R_j$
and $D$ as $\gamma^{SINR}_{sj}=\frac{E\{\Omega_{sj}|h_{sj}|^2\}}{\sigma_{sj}^2+E\{Y_{sj}\}}$, $\gamma^{SINR}_{jd}=\frac{E\{\Omega_{jd}|h_{jd}|^2\}}{\sigma_{jd}^2+E\{Y_{jd}\}}$, respectively.

Using these assumptions, for variable-gain relaying, the
amplification factor is given by
\begin{equation}
\label{ri5} G_j =
\frac{1}{\sqrt{\Omega_{sj}|h_{sj}|^2+\sigma_{sj}^2+\sum_{i=1}^{I^{sj}}\Omega_{ij}|h_{ij}|^2}}.
\end{equation}
Using (\ref{ri5}) in (\ref{ri4}), the instantaneous end-to-end SINR of the $j$-th relaying link can be derived as
\begin{equation}
\label{ri6}
\Gamma_j = \frac{\Gamma_{sj}\Gamma_{jd}}{\Gamma_{sj}+\Gamma_{jd}+1}
\end{equation}
where
\begin{equation}
\label{ri7} \Gamma_{sj} =
\frac{\Omega_{sj}|h_{sj}|^2}{\sigma_{sj}^2+Y_{sj}}
\end{equation}
and
\begin{equation}
\label{ri8} \Gamma_{jd} =
\frac{\Omega_{jd}|h_{jd}|^2}{\sigma_{jd}^2+Y_{jd}}.
\end{equation}

In relay selection, the relay with the largest end-to-end instantaneous SINR is
selected. Thus, the outage probability for a threshold of
$\gamma_{th}$ is given by
\begin{equation}
\label{ri9}
P_o(\gamma_{th}) = Pr\{\max\{\Gamma_j\}<\gamma_{th}\},\;\;\;j=1,2,\cdots,J
\end{equation}
where $\gamma_{th}=2^{2 R}-1$ and $R$ is the transmission rate.
In the next sections, this outage probability is derived in
different cases.
\section{Random Interferers}
We first consider the case when the interferers have random number and
locations.
In this case, the randomness comes from the Nakagami-$m$ fading powers,
the number of interferers $I^{sj}$ and $I^{jd}$ as well as the distances $l_{ij}$
and $l_{vj}$. In the first subsection, the PDFs of $Y_{sj}$ and
$Y_{jd}$ are derived using the generalized Gamma approximation (GGA), as their closed-form expressions are difficult to obtain, if not impossible. Then, closed-form expressions of the PDF and cumulative distribution function (CDF)
for $\Gamma_{sj}$ and $\Gamma_{jd}$ are derived. In
the third subsection, outage probability is
derived when the relay with the largest instantaneous end-to-end SINR is
selected. Note that the analysis of outage probability in this section can also be considered as an exact result when each relay and the destination have only one fixed interferer following generalized Gamma distribution.

\subsection{PDF of $Y_{sj}$ and $Y_{jd}$}
Exact closed-form expressions for the PDFs of $Y_{sj}$ and $Y_{jd}$ are
not available and are difficult to obtain. As a result, only moment generating functions (MGFs) of $Y_{sj}$ and $Y_{jd}$ for independent and identically distributed (i.i.d.)
channels are available in the literature
\cite{6516171}, which use either two- or three-dimensional integrals and thus are very complicated and not convenient to use.
Therefore, it is nearly impossible to get the closed-form expressions for the PDFs and CDFs of $Y_{sj}$ and $Y_{jd}$ from these MGFs because the inverse Laplace transform is further needed.
Thus, in the following, we will use GGA
by matching the first-order, second-order and third-order moments
of $Y_{sj}$ and $Y_{jd}$ to the first-order, second-order and
third-order moments of a generalized Gamma random variable.
To the best of the authors’ knowledge, none of the
works in the literature have considered using GGA to approximate the distribution of random interferers.
Numerical results in Section
\uppercase\expandafter{\romannumeral5} will show that the GGA approximation has a very good match with the simulation results.
As $Y_{sj}$ and $Y_{jd}$ have the same distribution but with different parameters, we approximate the distribution of $Y_{sj}$ first. One can get the approximate PDF of $Y_{sj}$ using GGA as
\begin{equation}\label{ri10}
\begin{aligned}
f_{Y_{sj}}(x;a_{sj},d_{sj},p_{sj})=\frac{p_{sj}\; a_{sj}^{-d_{sj}} x^{d_{sj}-1}
e^{-\left(\frac{x}{a_{sj}}\right)^{p_{sj}}}}{\Gamma
\left(\frac{d_{sj}}{p_{sj}}\right)}, x>0
\end{aligned}
\end{equation}
where $\Gamma(\cdot)$ is the Gamma function, $d_{sj}>0,p_{sj}>0$ are shape parameters and $a_{sj}>0$ is the scale
parameter to be determined \cite{6096455}. It is shown in Appendix A that one can calculate
$p_{sj}$ and $d_{sj}$ in (\ref{ri10}) by solving the two equations as
\begin{equation}\label{ri11}
\begin{aligned}
\left\{\begin{matrix}
&B\left(\frac{d_{sj}+1}{p_{sj}},\frac{d_{sj}+2}{p_{sj}}\right)=\frac{E\{Y_{sj}\}
E\{Y_{sj}^2\}
}{E\{Y_{sj}^3\}}B\left(\frac{d_{sj}}{p_{sj}},\frac{d_{sj}+3}{p_{sj}}\right)\\&B\left(\frac{d_{sj}+1}{p_{sj}},
\frac{d_{sj}+1}{p_{sj}}\right)=\frac{E\{Y_{sj}\}^2
}{E\{Y_{sj}^2\}}B\left(\frac{d_{sj}}{p_{sj}},\frac{d_{sj}+2}{p_{sj}}\right)
\end{matrix}\right.
\end{aligned}
\end{equation}
where $E\{Y_{sj}\}$, $E\{Y_{sj}^2\}$ and $E\{Y_{sj}^3\}$ can be found in (\ref{ri36}), (\ref{ri37}) and (\ref{ri38}), respectively, and then one can calculate $a_{sj}$ by inserting the solved values of $p_{sj}$ and $d_{sj}$ into one of the equations in (\ref{ri39}). Also, when all the interferers are i.i.d., with the help of $e^{\lambda}=\sum _{I=0}^\infty \frac{\lambda^I}{I!}$, $e^{\lambda}=\frac{1}{\lambda} \sum _{I=1}^\infty I \frac{\lambda^I}{I!}$ or $e^{\lambda}=\frac{1}{\lambda^2} \sum _{I=1}^\infty I (I-1) \frac{\lambda^I}{I!}$ \cite{Unnikrishna},
$E\{Y_{sj}\}$, $E\{Y_{sj}^2\}$ and $E\{Y_{sj}^3\}$ can be simplified as
\begin{eqnarray}  \label{ri12}     
\left\{                  
\begin{array}{lll}     
E\{Y_{sj}\}=I^{sj} K_{ij}P_{ij} \eta_{1,ij} \\    
E\{Y_{sj}^2\}=I^{sj} K^2_{ij}P^2_{ij}
\eta_{2,ij}\frac{m_{ij}+1}{m_{ij}} +
(I^{sj})^2
K_{i_1j}P_{i_1j}K_{i_2j}P_{i_2j}
\eta^2_{1,ij}\\E\{Y_{sj}^3\}=I^{sj}
K^3_{ij}P^3_{ij}\frac{(m_{ij}+1)(m_{ij}+2)}{m_{ij}^2}
\eta_{3,ij} +
(I^{sj})^2
K^2_{i_1j}P^2_{i_1j}K_{i_2j}P_{i_2j} \frac{m_{ij}+1}{m_{ij}}
\eta_{2,ij}
\eta_{1,ij} \\      
+ (I^{sj})^3
K_{i_1j}P_{i_1j}K_{i_2j}P_{i_2j}K_{i_3j}P_{i_3j}
\eta^3_{1,ij}.
\end{array}           
\right.              
\end{eqnarray}        

The PDF expression $f_{Y_{jd}}(x;a_{jd},d_{jd},p_{jd})$ of $Y_{jd}$ can be obtained using the same method as above. They are not listed here due to the limited space.

\subsection{PDF and CDF of $\Gamma_{sj}$ and $\Gamma_{jd}$}
Denote $W_{sj}=\Omega_{sj}\left | h_{sj} \right|^2$. Since $\left |
h_{sj} \right|^2$ is a Gamma random variable with shape
parameter $m_{sj}$ and scale parameter $1/m_{sj}$, $W_{sj}$ is also a
Gamma random variable with PDF
\begin{equation}\label{ri13}
\begin{aligned}
f_{W_{sj}}(x)=\left(\frac{m_{sj}}{\Omega_{sj}}\right)^{m_{sj}}\frac{x^{m_{sj}-1}}
{\Gamma (m_{sj})}e^{-\frac{m_{sj}}{\Omega_{sj}}x}, x>0.
\end{aligned}
\end{equation}
Also, denote $Z_{sj}=Y_{sj}+\sigma_{sj}^2$, where the PDF of $Z_{sj}$ is
determined by $f_{Z_{sj}}(x)=f_{Y_{sj}}(x-\sigma_{sj}^2)$.
Thus, one has
$\Gamma_{sj}=\frac{W_{sj}}{Z_{sj}}$ and the PDF of $\Gamma_{sj}$ is
given by
\begin{equation}\label{ri14}
\begin{aligned}
f_{\Gamma_{sj}}(u)=\int_{-\infty}^{\infty}\left | x \right
|f_{W_{sj}}\left ( x u \right )f_{Y_{sj}}\left ( x-\sigma_{sj}^2  \right
)dx.
\end{aligned}
\end{equation}
Using (\ref{ri10}) and (\ref{ri13}) in (\ref{ri14}) and defining
$p_{sj}=l_{sj}/k_{sj}$ such that $gcd(l_{sj},k_{sj})=1$ where $gcd(\cdot,\cdot)$ is
the great common devisor operator \cite{Gradshteyn}, one has
\begin{equation}\label{ri15}
\begin{aligned}
&f_{\Gamma_{sj}}(u)= \sum _{r_1=0}^{m_{sj}} \mu_{1,sj,r_1} e^{-\frac{m_{sj} \sigma_{sj}^2 u}{\Omega_{sj}}} u^{-d_{sj}+m_{sj}-r_1-1}
G_{l_{sj},k_{sj}}^{k_{sj},l_{sj}}\left(\mu_{0,sj} u^{-l_{sj}}|
\begin{array}{c}
I (l_{sj}, 1 - d_{sj} - r_1)\\ I (k_{sj}, 0) \\
\end{array}
 \right),
\end{aligned}
\end{equation}
where $\mu_{1,sj,r_1}= \frac{ l_{sj}^{d_{sj}+r_1-\frac{1}{2}} \sigma_{sj}^{2(m_{sj}-r_1)} \left(\frac{m_{sj}}{\Omega_{sj}}\right)^{-d_{sj}+m_{sj}-r_1}\sqrt{k_{sj}} p_{sj} m_{sj}! a_{sj}^{-d_{sj}} (2 \pi )^{-\frac{k_{sj}}{2}-\frac{l_{sj}}{2}+1}}{r_1! (m_{sj}-r_1)!\Gamma (m_{sj}) \Gamma \left(\frac{d_{sj}}{p_{sj}}\right)}$,
$\mu_{0,sj} = \\ k_{sj}^{-k_{sj}} l_{sj}^{l_{sj}} a_{sj}^{-k_{sj} p_{sj}} \left(\frac{m_{sj} }{\Omega_{sj}}\right)^{-l_{sj}}$,
$I(n,\xi)= \left(\xi/n,(\xi+1)/n,\cdots,(\xi+n-1)/n\right)$ and
$G_{a,b}^{c,d}(\cdot)$ denotes the Meijer'G-function
\cite{Gradshteyn} which is available as a built-in function in many
mathematical software packages, such as MATLAB, MATHEMATICA and
MAPLE.

\noindent \textit{Proof}: See Appendix B.

\noindent Also, one can get the CDF of $\Gamma_{sj}$ as
\begin{equation}\label{ri16}
\begin{aligned}
&F_{\Gamma_{sj}}(u)=1-\sum _{r_2=0}^{m_{sj}-1} \sum _{r_3=0}^{r_2}\mu_{2,sj,r_2,r_3}e^{-\frac{m_{sj} \sigma_{sj}^2 u}{\Omega_{sj}}}
u^{-d_{sj}+r_2-r_3}
G_{l_{sj},k_{sj}}^{k_{sj},l_{sj}}\left(\mu_{0,sj} u^{-l_{sj}}|
\begin{array}{c}
I (l_{sj}, 1 - d_{sj} - r_3)\\ I (k_{sj}, 0) \\
\end{array}
 \right)
\end{aligned}
\end{equation}
where $\mu_{2,sj,r_2,r_3}=\frac{l_{sj}^{d_{sj}+r_3-\frac{1}{2}} \sigma_{sj}^{2(r_2-r_3)} \left(\frac{m_{sj}}{\Omega_{sj}}\right)^{-d_{sj}+r_2-r_3} \sqrt{k_{sj}} p_{sj} a_{sj}^{-d_{sj}} (2 \pi )^{-\frac{k_{sj}}{2}-\frac{l_{sj}}{2}+1} }{r_3! (r_2-r_3)!\Gamma \left(\frac{d_{sj}}{p_{sj}}\right)}$.

\noindent \textit{Proof}: See Appendix C.

The PDF and CDF of $\Gamma_{sj}$ in (\ref{ri15}) and (\ref{ri16}) in terms of the Meijer'G-function are computationally complex. Therefore, we provide the high SINR approximations next that have simpler forms.
Using the following Taylor's series expansion
\begin{equation}\label{ri43}
\begin{aligned}
e^x=\sum _{n=0}^{N}\frac{x^n}{n!}+o(x^N),\;\;\;\;\; \text{as}\;\;\; x\rightarrow 0,
\end{aligned}
\end{equation}
where $o(x)$ denotes the higher-order term of an arbitrary function $a(x)$,
one can get the PDF and CDF of $\Gamma_{sj}$ as (\ref{wri17}) and (\ref{wri18}), respectively.
\begin{equation}\label{wri17}
\begin{aligned}
&f_{\Gamma_{sj}}(u)= \sum _{n_1=0}^{N_1}\sum _{n_3=0}^{N_3}\sum _{r_1=0}^{m_{sj}}\mu_{5,sj,r_1,n_1,n_3}u^{m_{sj}+n_1+n_3-1}
+o\left[(u/\Omega_{sj})^{N_1}\right]+o\left[(\sigma_{sj}^2 u/\Omega_{sj})^{N_3}\right],
\end{aligned}
\end{equation}
where $\mu_{5,sj,r_1,n_1,n_3}=\frac{(-1)^{n_1+n_3} a_{sj}^{n_1+r_1}m_{sj}^{m_{sj}+n_1+n_3+1} \Omega_{sj}^{-m_{sj}-n_1-n_3} \sigma_{sj}^{2(m_{sj}-r_1+n_3)} \Gamma \left(\frac{d_{sj}+n_1+r_1}{p_{sj}}\right)}{n_1! n_3! \Gamma (r_1+1) \Gamma (m_{sj}-r_1+1)\Gamma \left(\frac{d_{sj}}{p_{sj}}\right)}$,
\begin{equation}\label{wri18}
\begin{aligned}
&F_{\Gamma_{sj}}(u)=1-\sum _{n_2=0}^{N_2}\sum _{n_4=0}^{N_4}\sum _{r_2=0}^{m_{sj}-1}\sum _{r_3=0}^{r_2}\mu_{6,sj,r_2,r_3,n_2,n_4}u^{n_2+r_2+n_4} +o\left[(u/\Omega_{sj})^{N_2}\right]+o\left[(\sigma_{sj}^2 u/\Omega_{sj})^{N_4}\right],
\end{aligned}
\end{equation}
where $\mu_{6,sj,r_2,r_3,n_2,n_4}=\frac{(-1)^{n_2+n_4} a_{sj}^{n_2+r_3} \sigma_{sj}^{2(r_2-r_3+n_4)} \left(\frac{m_{sj}}{\Omega_{sj}}\right)^{n_2+r_2+n_4} \Gamma \left(\frac{d_{sj}+n_2+r_3}{p_{sj}}\right)}{n_2! n_4! r_3! (r_2-r_3)! \Gamma \left(\frac{d_{sj}}{p_{sj}}\right)}$.

\noindent \textit{Proof}: See Appendix D.

\noindent In high SINR conditions when $\Omega_{sj}\rightarrow \infty $ or $\sigma_{sj}^2\rightarrow 0$ or in the low outage regime when $u\rightarrow 0$, one can get rid of $o\left[(u/\Omega_{sj})^{N_1}\right]$, $o\left[(u/\Omega_{sj})^{N_2}\right]$, $o\left[(\sigma_{sj}^2 u/\Omega_{sj})^{N_3}\right]$ and $o\left[(\sigma_{sj}^2 u/\Omega_{sj})^{N_4}\right]$ in (\ref{wri17}) and (\ref{wri18}) to obtain corresponding approximations, respectively.

The PDF and CDF expressions $f_{\Gamma_{jd}}(u)$ and $F_{\Gamma_{jd}}(u)$ and their high SINR approximations can be obtained using the same method as above. They are not listed here to make the paper compact.

\subsection{Outage probability}
Using the derived PDF and CDFs of $\Gamma_{sj}$ and $\Gamma_{jd}$ in the previous subsection,
the CDF of the instantaneous end-to-end SINR in
(\ref{ri6}) can be derived as \cite{5876342}
\begin{equation}\label{ri19}
\begin{aligned}
F^E_{\Gamma_j}(x)&=\int_{0}^{\infty}Pr\left \{
\frac{\Gamma_{sj}\Gamma_{jd}}{\Gamma_{sj}+\Gamma_{jd}+1} \leq
x|t\right
\}f_{\Gamma_{sj}}(t)dt\\&=F_{\Gamma_{sj}}(x)+\int_{0}^{\infty}F_{\Gamma_{jd}}\left
( \frac{x^2+x+xt}{t} \right )f_{\Gamma_{sj}}(t+x)dt.
\end{aligned}
\end{equation}
One can see that (\ref{ri19}) only has one-dimensional integral, which can be calculated numerically using mathematical software. Also,
by using the lower bound as
\begin{equation}\label{ri20}
\begin{aligned}
&F_{\Gamma_j}(x)\geqslant 1-\left[1-F_{\Gamma_{sj}}(x)\right]\left[1-F_{\Gamma_{jd}}(x)\right]=F^{LB}_{\Gamma_j}(x),
\end{aligned}
\end{equation}
one can get the lower bound to the CDF as
\begin{equation}\label{ri21}
\begin{aligned}
&F^{LB}_{\Gamma_j}(x)=1-\sum _{r_2=0}^{m_{sj}-1} \sum _{r_3=0}^{r_2}\sum _{r'_2=0}^{m_{jd}-1} \sum _{r'_3=0}^{r'_2}\mu_{2,sj,r_2,r_3}\mu_{2,jd,r'_2,r'_3}e^{-\frac{m_{sj} \sigma_{sj}^2 x}{\Omega_{sj}}-\frac{m_{jd}\sigma_{jd}^2 x}{\Omega_{jd}}}
x^{-d_{sj}+r_2-r_3-d_{jd}+r'_2-r'_3}
\\& G_{l_{sj},k_{sj}}^{k_{sj},l_{sj}}\left(\mu_{0,sj} x^{-l_{sj}}|
\begin{array}{c}
I (l_{sj}, 1 - d_{sj} - r_3)\\ I (k_{sj}, 0) \\
\end{array}
 \right)
G_{l_{jd},k_{jd}}^{k_{jd},l_{jd}}\left(\mu_{0,jd} x^{-l_{jd}}|
\begin{array}{c}
I (l_{jd}, 1 - d_{jd} - r'_3)\\ I (k_{jd}, 0) \\
\end{array}
 \right),
\end{aligned}
\end{equation}
where $\mu_{2,jd,r'_2,r'_3}= \frac{l_{jd}^{d_{jd}+r'_3-\frac{1}{2}} \sigma_{jd}^{2(r'_2-r'_3)} \left(\frac{m_{jd}}{\Omega_{jd}}\right)^{-d_{jd}+r'_2-r'_3} \sqrt{k_{jd}} p_{jd} a_{jd}^{-d_{jd}} (2 \pi )^{-\frac{k_{jd}}{2}-\frac{l_{jd}}{2}+1} }{r'_3! (r'_2-r'_3)!\Gamma \left(\frac{d_{jd}}{p_{jd}}\right)}$ and
$\mu_{0,jd}=\\k_{jd}^{-k_{jd}} l_{jd}^{l_{jd}} a_{jd}^{-k_{jd} p_{jd}} \left(\frac{m_{jd} }{\Omega_{jd}}\right)^{-l_{jd}}$.

Also, if one inserts the CDFs of high SINR approximations into (\ref{ri20}) and with the help of (\ref{ri35}), one can get the asymptotic expression of the CDF as
\begin{equation}\label{ri22}
\begin{aligned}
&F^\infty_{\Gamma_{j}}(x)=1-\sum _{n_2=0}^{N_2}\sum _{n_4=0}^{N_4}\sum _{r_2=0}^{m_{sj}-1}\sum _{r_3=0}^{r_2} \sum _{n'_2=0}^{N'_2}\sum _{n'_4=0}^{N'_4}\sum _{r'_2=0}^{m_{jd}-1}\sum_{r'_3=0}^{r'_2}x^{n_2+r_2+n_4+n'_2+r'_2+n'_4}\\&
\frac{(-1)^{n_2+n_4+ n'_2+n'_4} \sigma_{sj}^{2(r_2-r_3+n_4)} \sigma_{jd}^{2(r'_2-r'_3+n'_4)} \left(\frac{m_{sj}}{\Omega_{sj}}\right)^{n_2+r_2+n_4} \left(\frac{m_{jd}}{\Omega_{jd}}\right)^{n'_2+r'_2+n'_4} E\left( Y_{sj}^{n_2+r_3}\right)E\left( Y_{jd}^{n'_2+r'_3}\right)}{n_2! n_4! r_3! (r_2-r_3)! n'_2! n'_4! r'_3! (r'_2-r'_3)! }
\end{aligned}
\end{equation}
where $E\left( Y_{sj}^{n_2+r_3}\right)=\frac{a_{sj}^{n_2+r_3}\Gamma \left(\frac{d_{sj}+n_2+r_3}{p_{sj}}\right)}{\Gamma \left(\frac{d_{sj}}{p_{sj}}\right)}$,
$E\left( Y_{jd}^{n'_2+r'_3}\right)=\frac{a_{jd}^{n'_2+r'_3}\Gamma \left(\frac{d_{jd}+n'_2+r'_3}{p_{jd}}\right)}{\Gamma \left(\frac{d_{jd}}{p_{jd}}\right)}$.

Then, the outage probability for AF relay selection is given by
\begin{equation}\label{www2}
\begin{aligned}
P^\Psi _o(\gamma_{th})=\prod_{j=1}^J F^\Psi _{\Gamma_j}(\gamma_{th})
\end{aligned}
\end{equation}
where $\Psi$ in (\ref{www2}) can be $E$ using $F^E_{\Gamma_j}(x)$ to get the exact outage probability, $LB$ using $F^{LB}_{\Gamma_j}(x)$ to get lower bound and $\infty$ using $F^{\infty}_{\Gamma_j}(\gamma_{th})$ to get the asymptotic expression for the outage probability.

Using the simple form in (\ref{ri22}), several insights can be obtained. For example, in the low outage regime when $\gamma_{th}\rightarrow 0$, the above result becomes exact. Also, one can see from (\ref{ri22}) that when the Nakagami-$m$ parameters of the interference $m_{ij}$ or $m_{vj}$ ($\gg$ 1) are large and increase, the outage probability remains almost unchanged. This is because $m_{ij}$ or $m_{vj}$ only have an influence on the order of interference power, as can be seen from (\ref{ri36}), (\ref{ri37}) and (\ref{ri38}) in Appendix A, where the $c$-th order moment of interference power $E\left( Y_{sj}\right)$ or $E\left( Y_{jd}\right)$ remains almost unchanged when $m_{ij}$ or $m_{vj}$ are large. For small values of $m_{ij}$ or $m_{vj}$ ($\approx 1$), they still have some influence on the outage probability, as $\frac{m_{ij}+1}{m_{ij}}$ in (\ref{ri37}) and $\frac{(m_{ij}+1)(m_{ij}+2)}{m^2_{ij}}$ in (\ref{ri38}) cannot be ignored if $m_{ij}\approx 1$ (They approach 1 when $m_{ij}\rightarrow 0$). This phenomenon will be shown in Fig.\;\ref{mij} by simulation in Section \uppercase\expandafter{\romannumeral5}.

Also, one can see from (\ref{ri22}) that, when the SINR increases (i.e. $\Omega_{sj}$ or $\Omega_{jd}$ increase, or $\sigma^2_{sj}$ or $\sigma^2_{jd}$ decrease, or $E\left( Y_{sj}\right)$ or $E\left( Y_{jd}\right)$ decrease), the outage probability decreases accordingly, which will be shown in Fig.\;\ref{SNR=15nINR=20} and Fig.\;\ref{SNR=20nINR=20} in Section \uppercase\expandafter{\romannumeral5}.
Further, for fixed SINR, if one decreases INR (i.e. decreases the interference power $E\left( Y_{sj}\right)$ or $E\left( Y_{jd}\right)$), the outage probability still decreases. This is
because the order of the interference power $E\left( Y_{sj}^{n_2+r_3}\right)$ or $E\left( Y_{jd}^{n'_2+r'_3}\right)$ in (\ref{ri22}) also affects the outage probability. The rate of decrease becomes small when the diversity order (determined by $m_{sj}$ or $m_{jd}$) is small. Therefore, when the signal experiences Rayleigh fading (i.e. $m_{sj}=1$ or $m_{jd}=1$), the outage probability remains almost unchanged if one changes INR but keeps SINR fixed. These explanations will be verified
in Fig.\;\ref{SNR=15nINR=0} and Fig.\;\ref{SNR=15nINR=20} in Section \uppercase\expandafter{\romannumeral5}.

Also, since the possible boundary of the interferers and the pass loss between the interferer and the signal have influence on the order of interference power (referring to (\ref{ri36}), (\ref{ri37}) and (\ref{ri38}) in Appendix A), changing the possible boundary and the pass loss also changes the outage probability, even if one keeps the SINR and INR fixed.
Similar to before, for the Rayleigh case, the outage probability changes little if one keeps SINR and INR fixed.
These discussions will be verified in Fig.\;\ref{theta5} and Fig.\;\ref{L20} in Section \uppercase\expandafter{\romannumeral5}.

Another insight that can be obtained from (\ref{ri22}) is that, when $m_{sj}$ or $m_{jd}$ increase, the outage probability decreases, as $m_{sj}$ or $m_{jd}$ are in the upper limits of the summations in (\ref{ri22}), which will also be examined in Fig.\;\ref{SNR=15nINR=0} - Fig.\;\ref{L20} in Section \uppercase\expandafter{\romannumeral5}.

\subsection{Special cases}
For the sake of simplicity, we now focus on the case when the interferences are dominant at both relay and destination. By setting $\sigma_{sj}^2\approx 0$ and $\sigma_{jd}^2\approx 0$ in (\ref{ri22}), one can get the asymptotic CDF in this case as
\begin{equation}\label{wri22}
\begin{aligned}
&F^\infty_{\Gamma_{j}}(x)=1-\sum _{n_2=0}^{N_2}\sum _{r_2=0}^{m_{sj}-1}\sum _{n'_2=0}^{N'_2}\sum _{r'_2=0}^{m_{jd}-1}\frac{(-1)^{n_2+n'_2} \left(\frac{m_{sj}}{\Omega_{sj}}\right)^{n_2+r_2} \left(\frac{m_{jd}}{\Omega_{jd}}\right)^{n'_2+r'_2} E\left( Y_{sj}^{n_2+r_2}\right)E\left( Y_{jd}^{n'_2+r'_2}\right)
}{n_2! r_2!n'_2! r'_2!}\\&x^{n_2+r_2+n'_2+r'_2},
\end{aligned}
\end{equation}
where $E\left( Y_{sj}^{n_2+r_2}\right)=\frac{a_{sj}^{n_2+r_2}\Gamma \left(\frac{d_{sj}+n_2+r_2}{p_{sj}}\right)}{\Gamma \left(\frac{d_{sj}}{p_{sj}}\right)}$,
$E\left( Y_{jd}^{n'_2+r'_2}\right)=\frac{a_{jd}^{n'_2+r'_2}\Gamma \left(\frac{d_{jd}+n'_2+r'_2}{p_{jd}}\right)}{\Gamma \left(\frac{d_{jd}}{p_{jd}}\right)}$.

In another special case when the signal experiences Rayleigh fading channel, by setting $m_{sj}= 1$ and $m_{jd}=1$, (\ref{ri22}) is specialized to
\begin{equation}\label{wwri22}
\begin{aligned}
&F^\infty_{\Gamma_{j}}(x)=1-\sum _{n_2=0}^{N_2}\sum _{n_4=0}^{N_4}\sum _{n'_2=0}^{N'_2}\sum _{n'_4=0}^{N'_4}
\frac{(-1)^{n_2+n_4+n'_2+n'_4} \Omega_{sj}^{-n_2-n_4} \Omega_{jd}^{-n'_2-n'_4}\sigma_{sj}^{2 n_4} \sigma_{jd}^{2 n'_4} E\left( Y_{sj}^{n_2}\right)E\left( Y_{jd}^{n'_2}\right)}{n_2! n_4! n'_2! n'_4!}\\&x^{n_2+n_4+n'_2+n'_4},
\end{aligned}
\end{equation}
where $E\left( Y_{sj}^{n_2}\right)=\frac{a_{sj}^{n_2}\Gamma \left(\frac{d_{sj}+n_2}{p_{sj}}\right)}{\Gamma \left(\frac{d_{sj}}{p_{sj}}\right)}$,
$E\left( Y_{jd}^{n'_2}\right)=\frac{a_{jd}^{n'_2}\Gamma \left(\frac{d_{jd}+n'_2}{p_{jd}}\right)}{\Gamma \left(\frac{d_{jd}}{p_{jd}}\right)}$.

Then, using (\ref{www2}), the outage probability for the two special cases above can be obtained.

\section{Fixed Interferers}
\label{three}
In this section, we consider the case when the interferers have fixed number and locations. In this case, the numbers of interferers $I^{sj}$ and $I^{jd}$ as well as the distances $l_{ij}$ and $l_{vj}$ are deterministic such that they can be all treated as
constants. The only randomness comes from the Nakagami-$m$ fading.
Thus, $\Gamma_{sj}$ is a function of only the random channel gains.
Note that similar derivations have also been conducted for dual-hop AF relaying without relay selection in the literature \cite{5426509,5545638,5740505,6324367,6189013,6130607,6209368,5992706,5611617,5784302}.
However, they either consider interferences at only one of the relay and the destination \cite{5426509,5545638,5740505,6189013,5784302},
only a single interferer \cite{6130607}, for fixed-gain relaying \cite{6324367}, for performance upper bounds \cite{6209368}, for interference-limited case with identical Nakagami-$m$ channels \cite{5992706}, or for Rayleigh channels \cite{5611617}. In the following, we will derive the exact performance for the case when
both the relay and the destination suffer from multiple
non-identically distributed Nakagami-$m$ interferers and we also consider relay selection in our derivation.


\subsection{PDF and CDF of $\Gamma_{sj}$ and $\Gamma_{jd}$}
Similarly, we derive the PDF and CDF of $\Gamma_{sj}$ first. Since $Y_{sj}=\sum_{i=1}^{I^{sj}}\Omega_{ij}|h_{ij}|^2$ and $|h_{ij}|^2$ are independent Gamma random variables, by proper scaling, $Y_{sj}$ is actually a sum of independent Gamma random variables. A closed-form expression for the PDF of this sum was derived in \cite{Mosch}. However, this expression uses an infinite series in order to consider the general case of arbitrary shape parameters and scale parameters. To avoid this infinite series in our result, we use the PDF in \cite{Mathai} for the case when the Nakagami-$m$ parameters could be different but are integers. In this case, the PDF of $Y_{sj}$ is given by \cite[eq. (4)]{Mathai}
\begin{equation}
\label{ri24}
f_{Y_{sj}}(x) = \left[\prod_{i^*=1}^{I^{sj}}\left(-\frac{\Omega_{i^*j}}{m_{i^*j}}\right)^{-m_{i^*j}}\right]\sum_{i=1}^{I^{sj}}\sum_{r=1}^{m_{ij}}\frac{(-1)^rb_{ir}}{(r-1)!}x^{r-1}e^{-\frac{m_{ij}}{\Omega_{ij}}x}, x>0
\end{equation}
where $b_{ir}$ is a constant given by \cite[eq. (5)]{Mathai}, $\Omega_{i^*j}$, $m_{i^*j}$ are the same as $\Omega_{ij}$, $m_{ij}$, respectively for the same $i$ and $j$
and all other symbols are defined as before.
It is derived in Appendix E that the PDF of $\Gamma_{sj}$ can be written as
\begin{eqnarray}
\label{ri25}
f_{\Gamma_{sj}}(u)&=&
\sum_{i=1}^{I^{sj}}\sum_{r=1}^{m_{ij}}\sum_{f=0}^{m_{sj}}\varphi_{1,sj,ij,i,r,f}
\frac{u^{m_{sj}-1}e^{-\frac{m_{sj}}{\Omega_{sj}}\sigma_{sj}^2u}}
{(\frac{m_{sj}}{\Omega_{sj}}u+\frac{m_{ij}}{\Omega_{ij}})^{f+r}},
\end{eqnarray}
where $\varphi_{1,sj,ij,i,r,f}=\left[\prod_{i^*=1}^{I^{sj}}
\left(-\frac{\Omega_{i^*j}}{m_{i^*j}}\right)^{-m_{i^*j}}\right]
\frac{(-1)^rb_{ir}\left(\frac{m_{sj}}
{\Omega_{sj}}\right)^{m_{sj}}\left(_f^{m_{sj}}\right)
(\sigma_{sj}^2)^{m_{sj}-f}\Gamma(f+r)}{\Gamma(m_{sj})(r-1)!}$,

\noindent and the CDF of $\Gamma_{sj}$ can be derived as
\begin{eqnarray}
\label{ri26}
F_{\Gamma_{sj}}(u)&=&1-\sum_{i=1}^{I^{sj}}
\sum_{r=1}^{m_{ij}}\sum_{f=0}^{m_{sj}-1}\sum_{h=0}^f\varphi_{2,sj,ij,i,r,f,h}
\frac{u^f e^{-\frac{m_{sj}}{\Omega_{sj}}\sigma_{sj}^2u}}
{(\frac{m_{sj}}{\Omega_{sj}}u+\frac{m_{ij}}{\Omega_{ij}})^{h+r}},
\end{eqnarray}
where $\varphi_{2,sj,ij,i,r,f,h}=\left[\prod_{i^*=1}^{I^{sj}}
\left(-\frac{\Omega_{i^*j}}{m_{i^*j}}\right)^{-m_{i^*j}}\right]
\frac{(-1)^r b_{ir}\left(\frac{m_{sj}}{\Omega_{sj}}\right)^f \left(_h^f\right)
(\sigma_{sj}^2)^{f-h}\Gamma(h+r)}{(r-1)!f!}$.

\noindent Using (\ref{ri43}) and the following Taylor's series expansion
\begin{equation}\label{wri43}
\begin{aligned}
(1+x)^{-n}=\sum _{i=0}^{N}\left(_i^{-n}\right)x^i+o(x^N),\;\;\;\;\; \text{as}\;\;\; x\rightarrow 0,
\end{aligned}
\end{equation}
where $n$ and $N$ are positive integers, one can get the high SINR approximations for PDF and CDF of $\Gamma_{sj}$ as (\ref{wri25}) and (\ref{wri26}), respectively.
\begin{equation}\label{wri25}
\begin{aligned}
&f_{\Gamma_{sj}}(u)=
\sum_{i=1}^{I^{sj}}\sum_{n_5=0}^{N_5}\sum_{n_6=0}^{N_6}\sum_{r=1}^{m_{ij}}\sum_{f=0}^{m_{sj}}
\varphi_{3,sj,ij,i,r,f,n_5,n_6}
u^{m_{sj}+n_5+n_6-1}+o\left[(\sigma_{sj}^2 u/\Omega_{sj})^{N_5}\right]\\&+o\left[(u \Omega_{ij}/\Omega_{sj})^{N_6}\right]
\end{aligned}
\end{equation}
where $\varphi_{3,sj,ij,i,r,f,n_5,n_6}=\\\left[\prod_{i^*=1}^{I^{sj}}
\left(-\frac{\Omega_{i^*j}}{m_{i^*j}}\right)^{-m_{i^*j}}\right]
\frac{(-1)^{r+n_5}b_{ir}\left(\frac{m_{sj}}
{\Omega_{sj}}\right)^{m_{sj}+n_5+n_6}
\left(\frac{m_{ij}}{\Omega_{ij}}\right)^{-f-r-n_6}
\left(_f^{m_{sj}}\right)\left(_{n_6}^{-f-r}\right)
(\sigma_{sj}^2)^{m_{sj}-f+n_5}\Gamma(f+r)}{\Gamma(m_{sj})(r-1)!n_5!}$,
\begin{equation}\label{wri26}
\begin{aligned}
&F_{\Gamma_{sj}}(u)=1-\sum_{i=1}^{I^{sj}}\sum_{n_7=0}^{N_7}\sum_{n_8=0}^{N_8}
\sum_{r=1}^{m_{ij}}\sum_{f=0}^{m_{sj}-1}\sum_{h=0}^f\varphi_{4,sj,ij,i,r,f,h,n_7,n_8}
u^{f+n_7+n_8}+o\left[(\sigma_{sj}^2 u/\Omega_{sj})^{N_7}\right]\\&+o\left[(u \Omega_{ij}/\Omega_{sj})^{N_8}\right],
\end{aligned}
\end{equation}
where $\varphi_{4,sj,ij,i,r,f,h,n_7,n_8}=\\\left[\prod_{i^*=1}^{I^{sj}}
\left(-\frac{\Omega_{i^*j}}{m_{i^*j}}\right)^{-m_{i^*j}}\right]
\frac{(-1)^{r+n_7} b_{ir}\left(\frac{m_{sj}}{\Omega_{sj}}\right)^{f+n_7+n_8}
\left(\frac{m_{ij}}{\Omega_{ij}}\right)^{-h-r-n_8}
\left(_h^f\right)\left(_{n_8}^{-h-r}\right)
(\sigma_{sj}^2)^{f-h+n_7}\Gamma(h+r)}{(r-1)!f!n_7!}$.

\noindent In high SINR conditions, $o\left[(\sigma_{sj}^2 u/\Omega_{sj})^{N_5}\right]$, $o\left[(u \Omega_{ij}/\Omega_{sj})^{N_6}\right]$, $o\left[(\sigma_{sj}^2 u/\Omega_{sj})^{N_7}\right]$ and $o\left[(u \Omega_{ij}/\Omega_{sj})^{N_8}\right]$ in the above equations can be removed
to obtain corresponding approximations.

The PDF and CDF expressions $f_{\Gamma_{jd}}(u)$ and $F_{\Gamma_{jd}}(u)$ and their high SINR approximations can be also obtained using the same methods as above.
\subsection{Outage probability}
Using the derived exact PDF and CDFs of $\Gamma_{sj}$ and $\Gamma_{jd}$ into (\ref{ri19}), one can get the CDF of the instantaneous end-to-end SINR in one-dimensional integral, which can be calculated numerically using mathematical software.

Then, following the same process as in Section \uppercase\expandafter{\romannumeral3}, one can get the lower bound of the CDF as (\ref{ri28}), when the exact CDFs of $\Gamma_{sj}$ and $\Gamma_{jd}$ are used,
\begin{equation}\label{ri28}
\begin{aligned}
&F_{\Gamma_j}(x)\geqslant1-\sum_{i=1}^{I^{sj}}
\sum_{r=1}^{m_{ij}}\sum_{f=0}^{m_{sj}-1}\sum_{h=0}^f
\sum_{v=1}^{I^{jd}}
\sum_{r'=1}^{m_{vj}}\sum_{f'=0}^{m_{jd}-1}\sum_{h'=0}^{f'}\frac{u^{f} e^{-\frac{m_{sj}}{\Omega_{sj}}\sigma_{sj}^2u}}
{(\frac{m_{sj}}{\Omega_{sj}}u+\frac{m_{ij}}{\Omega_{ij}})^{h+r}}\frac{u^{f'} e^{-\frac{m_{jd}}{\Omega_{jd}}\sigma_{jd}^2u}}
{(\frac{m_{jd}}{\Omega_{jd}}u+\frac{m_{vj}}{\Omega_{vj}})^{h'+r'}}\\&\varphi_{2,sj,ij,i,r,f,h}
\varphi_{2,jd,vd,v,r',f',h'}=F^{LB}_{\Gamma_j}(x),
\end{aligned}
\end{equation}
where $\varphi_{2,jd,vd,v,r',f',h'}=\left[\prod_{v^*=1}^{I^{jd}}
\left(-\frac{\Omega_{v^*j}}{m_{v^*j}}\right)^{-m_{v^*j}}\right]
\frac{(-1)^{r'} b_{vr'}\left(\frac{m_{jd}}{\Omega_{jd}}\right)^{f'} \left(_{h'}^{f'}\right)
(\sigma_{jd}^2)^{f'-h'}\Gamma(h'+r')}{(r'-1)!f'!}$.

\noindent Also, using high SINR approximations of $F_{\Gamma_{sj}}$ and $F_{\Gamma_{jd}}$,
one can get the asymptotic expression of the CDF as
\begin{equation}\label{ee1}
\begin{aligned}
&F^\infty_{\Gamma_{j}}(x)=1-\sum_{i=1}^{I^{sj}}\sum_{n_7=0}^{N_7}\sum_{n_8=0}^{N_8}
\sum_{r=1}^{m_{ij}}\sum_{f=0}^{m_{sj}-1}\sum_{h=0}^f
\sum_{v=1}^{I^{jd}}\sum_{n'_7=0}^{N'_7}\sum_{n'_8=0}^{N'_8}
\sum_{r'=1}^{m_{vj}}\sum_{f'=0}^{m_{jd}-1}\sum_{h'=0}^{f'}u^{f+n_7+n_8+f'+n'_7+n'_8}\\&\varphi_{4,sj,ij,i,r,f,h,n_7,n_8}
\varphi_{4,jd,vd,v,r',f',h',n'_7,n'_8},
\end{aligned}
\end{equation}
where $\varphi_{4,jd,vd,v,r',f',h',n'_7,n'_8}=\\\left[\prod_{v^*=1}^{I^{jd}}
\left(-\frac{\Omega_{v^*j}}{m_{v^*j}}\right)^{-m_{v^*j}}\right]
\frac{(-1)^{r'+n'_7} b_{vr'}\left(\frac{m_{jd}}{\Omega_{jd}}\right)^{f'+n'_7+n'_8}
\left(\frac{m_{vj}}{\Omega_{vj}}\right)^{-h'-r'-n'_8}
\left(_{h'}^{f'}\right)\left(_{n'_8}^{-h'-r'}\right)
(\sigma_{jd}^2)^{f'-h'+n'_7}\Gamma(h'+r')}{(r'-1)!f'!n'_7!}$.

Then, using (\ref{www2}), the outage probability can be obtained.

\subsection{When the interferences are dominant}

In the case when the interferences are dominant at both relay and destination such that $\sigma_{sj}^2\approx 0$ and $\sigma_{jd}^2\approx 0$, exact CDF expression (\ref{ri19}) can be solved in closed-form as
\begin{equation}\label{ri27}
\begin{aligned}
&F_{\Gamma_j}(x)=F_{\Gamma_{sj}}(x)+\sum_{i=1}^{I^{sj}}\sum_{r_1=1}^{m_{ij}}
\varphi_7\varphi_9 (x)- \sum_{i=1}^{I^{sj}}\sum_{v=1}^{I^{jd}}\sum_{r_1=1}^{m_{ij}}\sum_{r_2=1}^{m_{vj}}
\sum_{f=0}^{m_{jd}-1}\sum _{j_1=0}^{m_{sj}-1}\sum _{j_2=0}^{f}
\varphi_8
\varphi_{10} (x)
\end{aligned}
\end{equation}
where
$\varphi_7=\left(\frac{m_{sj}}{\Omega_{sj}}\right)^{-r_1}
\left[\prod_{i^*=1}^{I^{sj}}\left(-\frac{\Omega_{i^*j}}{m_{i^*j}}\right)^{-m_{i^*j}}\right]
\frac{(-1)^{r_1} b_{ir_1}\Gamma(m_{sj}+r_1)B(1,r_1)}{\Gamma(m_{sj})(r_1-1)!}$,\\
$\varphi_8=\left(\frac{m_{sj}}{\Omega_{sj}}\right)^{m_{sj}}
\left[\prod_{i^*=1}^{I^{sj}}\left(-\frac{\Omega_{i^*j}}{m_{i^*j}}\right)^{-m_{i^*j}}\right]
\left[\prod_{v^*=1}^{I^{jd}}\left(-\frac{\Omega_{v^*j}}{m_{v^*j}}
\right)^{-m_{v^*j}}\right]\left(_{j_2}^{f}\right)\left(_{j_1}^{m_{sj}-1}\right)\\
\frac{(-1)^{r_1+r_2}b_{ir_1}b_{vr_2}\Gamma(m_{sj}+r_1)\Gamma(f+r_2)\Omega_{jd}^{r_2} (\Omega_{ij} \Omega_{sj})^{m_{sj}+r_1} m_{jd}^{j_1+j_2+1} \Omega_{vj}^{j_1+j_2+r_2+1}}{(r_2-1)!f!\Gamma(m_{sj})(r_1-1)!}B(j_1+j_2+r_2+1,m_{sj}
+r_1-j_1-j_2+f-1)$,
$\varphi_9 (x)=\; _2F_1\left(1,1-m_{sj};r_1+1;-\frac{m_{ij} \Omega_{sj}}{m_{sj} \Omega_{ij}x}\right) \left(\frac{m_{ij} \Omega_{sj}}{m_{sj} \Omega_{ij}}+x\right)^{-m_{sj}-r_1+1}x^{m_{sj}-1}$,
$\varphi_{10} (x)=(x+1)^{j_1+1} x^{j_2+m_{sj}}(m_{ij} \Omega_{sj}+m_{sj} \Omega_{ij} x)^{-m_{sj}-r_1}(m_{jd} \Omega_{vj} x+m_{vj} \Omega_{jd})^{-j_1-j_2-r_2-1} \, _2F_1 (m_{sj}+r_1,j_1+j_2+r_2+1;m_{sj}+r_1+f+r_2;1-\frac{m_{jd} \Omega_{vj}m_{sj} \Omega_{ij} \left(x^2+x\right)}{\left(m_{ij} \Omega_{sj}+m_{sj} \Omega_{ij}x\right) (m_{vj} \Omega_{jd}+m_{jd} \Omega_{vj} x)})$
and $_2F_1(\cdot,\cdot;\cdot;\cdot)$ is the hypergeometric function.

\noindent \textit{Proof}: See Appendix F.

\noindent Note that (\ref{ri27}) is a very good closed-form approximation to the exact CDF when INR is large in the case of fixed interferences.

\subsection{When the interferences are i.i.d.}
In this subsection, we focus on the case when all the interferences are i.i.d.
One can see that the PDF $f_{Y_{sj}}(x)$ in (\ref{ri24}) is only suitable when the Nakagami-$m$ parameters of interferences are different. When all the interferences are i.i.d., (\ref{ri24}) simply becomes
\begin{equation}\label{ri30}
\begin{aligned}
f_{Y_{sj}}(x)=\left(\frac{m_{ij}}{\Omega_{ij}}\right)^{I^{sj}m_{ij}}\frac{x^{I^{sj} m_{ij}-1}}
{\Gamma (I^{sj}m_{ij})}e^{-\frac{m_{ij}}{\Omega_{ij}}x}, x>0
\end{aligned}
\end{equation}
where $b_{ir}$ does not exist any more and all other symbols are defined as before.
Using (\ref{ri13}), (\ref{ri14}), (\ref{ri20}) and (\ref{ri30}), the lower bound can be derived as
\begin{equation}\label{ew1}
\begin{aligned}
&F_{\Gamma_j}(x)\geqslant1-\varphi_{12}\sum _{v_1=0}^{m_{sj}-1} \sum _{v_2=0}^{m_{jd}-1} \sum _{s_1=0}^{v_1} \sum _{s_2=0}^{v_2}\varphi_{13}\varphi_{14}(x)=F^{LB}_{\Gamma_j}(x)
\end{aligned}
\end{equation}
where $\varphi_{12}=\frac{\left(\frac{\Omega_{vj}}{m_{vj}}\right)^{-I^{jd} m_{vj}} \left(\frac{\Omega_{ij}}{m_{ij}}\right)^{-I^{sj} m_{ij}} e^{-\frac{m_{jd} \sigma_{jd}^2 x}{\Omega_{jd}}-\frac{m_{sj} \sigma_{sj}^2 x}{\Omega_{sj}}}}{\Gamma (I^{jd} m_{vj}) \Gamma (I^{sj} m_{ij})}$, \\$\varphi_{13}=\frac{\left(\frac{\Omega_{jd}}{m_{jd}}\right)^{-v_2} \left(\frac{\Omega_{sj}}{m_{sj}}\right)^{-v_1} (\sigma_{jd}^2)^{v_2-s_2} (\sigma_{sj}^2)^{v_1-s_1} \Gamma (I^{jd} m_{vj}+s_2) \Gamma (I^{sj} m_{ij}+s_1)}{s_1! s_2! (v_1-s_1)! (v_2-s_2)!}$, \\ $\varphi_{14}(x)=x^{v_1+v_2} \left(\frac{m_{jd} x}{\Omega_{jd}}+\frac{m_{vj}}{\Omega_{vj}}\right)^{-I^{jd} m_{vj}-s_2} \left(\frac{m_{ij}}{\Omega_{ij}}+\frac{m_{sj} x}{\Omega_{sj}}\right)^{-I^{sj} m_{ij}-s_1}$.

\noindent Similar, the asymptotic outage probability in the low outage regime can be derived as
\begin{equation}\label{ew2}
\begin{aligned}
&F^\infty_{\Gamma_{j}}(x)=1-\sum _{v_1=0}^{m_{sj}-1} \sum _{v_2=0}^{m_{jd}-1} \sum _{s_1=0}^{v_1} \sum _{s_2=0}^{v_2} \sum _{n_9=0}^{N_9} \sum _{n_{10}=0}^{N_{10}} \sum _{n_{11}=0}^{N_{11}} \sum _{n_{12}=0}^{N_{12}}
\varphi_{15}
\varphi_{16}x^{n_{10}+n_{11}+n_{12}+n_9+v_1+v_2},
\end{aligned}
\end{equation}
where $\varphi_{15}=\frac{(-1)^{n_{10}+n_9} \Gamma (I^{jd} m_{vj}+s_2) \Gamma (I^{sj} m_{ij}+s_1) \left(
_{ n_{12} }^{-I^{jd} m_{vj}-s_2 }
\right) \left(
_{n_{11} }^{-I^{sj} m_{ij}-s_1}
\right)}{n_{10}! n_9! s_1! s_2! (v_1-s_1)! (v_2-s_2)! \Gamma (I^{jd} m_{vj}) \Gamma (I^{sj} m_{ij})}$, \\$\varphi_{16}=\left(\frac{\Omega_{ij}}{m_{ij}}\right)^{n_{11}+s_1} \left(\frac{\Omega_{vj}}{m_{vj}}\right)^{n_{12}+s_2} (\sigma_{jd}^2)^{n_{10}-s_2+v_2} (\sigma_{sj}^2)^{n_9-s_1+v_1} \left(\frac{\Omega_{jd}}{m_{jd}}\right)^{-n_{10}-n_{12}-v_2} \left(\frac{\Omega_{sj}}{m_{sj}}\right)^{-n_{11}-n_9-v_1}$.

Then, the lower bound and asymptotic expression for the outage probability can be derived by
using (\ref{ew1}) and (\ref{ew2}) in (\ref{www2}).
Several insights can be obtained from (\ref{ew2}) for i.i.d. fixed interferers. For example, one can see that with the increase of the Nakagami-$m$ parameters of the interference $m_{ij}$ or $m_{vj}$, the outage probability remains almost the same, which will be examined in Fig. \ref{mijmvj}.
Similar to the analysis for random interferers, with the increase of SINR, the outage probability for fixed interferer decreases accordingly, which will be examined in Fig.\;\ref{s15n20} and Fig.\;\ref{s20n20} in Section \uppercase\expandafter{\romannumeral5}. However, different from random interferers, changing INR will not have a great influence on the outage probability for fixed interferers if the SINR is fixed. This is because SINR dominates the outage probability for fixed interferes and changing INR just change the ratio between the noise power and the interference power but has negligible influence on the overall outage probability, which will be shown in Fig.\;\ref{s15n0} and Fig.\;\ref{s15n20} in Section \uppercase\expandafter{\romannumeral5}.
Another insight that can be obtained from (\ref{ew2}) is that, with the increase of $m_{sj}$ or $m_{jd}$, the outage probability decreases, as $m_{sj}$ or $m_{jd}$ are in the upper limits of the summations in (\ref{ew2}), which will be examined in Fig.\;\ref{s15n0} - Fig.\;\ref{s20n20} via simulation in Section \uppercase\expandafter{\romannumeral5}.

In the i.i.d. case when the interferences are dominant at both relay and destination, one can further get the closed-form expression of the CDF as
\begin{equation}\label{ri31}
\begin{aligned}
&F_{\Gamma_j}(x)=1-\sum _{f_1=0}^{m_{sj}-1}\varphi_{17}x^{f_1} \left(\frac{m_{ij}}{\Omega_{ij}}+\frac{m_{sj} x}{\Omega_{sj}}\right)^{-f_1-I^{sj} m_{ij}}+\varphi_{18}\varphi_{20}(x)
-\sum _{f_2=0}^{m_{jd}-1}\sum _{s_1=0}^{f_2}\sum _{s_2=0}^{m_{sj}-1}\varphi_{19}\varphi_{21}(x)
\end{aligned}
\end{equation}
where $\varphi_{17}=\frac{\Gamma (f_1+I^{sj} m_{ij})}{f_1! \left(\frac{\Omega_{sj}}{m_{sj}}\right)^{f_1} \left(\frac{\Omega_{ij}}{m_{ij}}\right)^{I^{sj} m_{ij}} \Gamma (I^{sj} m_{ij})}$, $\varphi_{18}=\frac{m_{sj}^{-I^{sj} m_{ij}} \left(\frac{\Omega_{ij}}{m_{ij}}\right)^{-I^{sj} m_{ij}} \Omega_{sj}^{I^{sj} m_{ij}} B(1,I^{sj} m_{ij})}{B(I^{sj} m_{ij},m_{sj})}$, $\varphi_{19}= \frac{\Gamma (f_2+I^{jd} m_{vj}) m_{jd}^{-I^{jd} m_{vj}} \Omega_{jd}^{I^{jd} m_{vj}} \left(\frac{\Omega_{vj}}{m_{vj}}\right)^{-I^{jd} m_{vj}} m_{sj}^{-I^{sj} m_{ij}} \left(\frac{\Omega_{ij}}{m_{ij}}\right)^{-I^{sj} m_{ij}} \Omega_{sj}^{I^{sj} m_{ij}} \Gamma (I^{sj} m_{ij}+m_{sj})}{\Gamma (I^{jd} m_{vj}) \Gamma (I^{sj} m_{ij})s_1! s_2! (f_2-s_1)! \Gamma (m_{sj}-s_2)}\\ B(I^{jd} m_{vj}+s_1+s_2+1, f_2+I^{sj} m_{ij}+m_{sj}-s_1-s_2-1)$,
$\varphi_{20}(x)=x^{m_{sj}-1} \left(\frac{\Omega_{sj}m_{ij}}{m_{sj} \Omega_{ij}}+x\right)^{-I^{sj} m_{ij}-m_{sj}+1} \\\, _2F_1\left(1,1-m_{sj};I^{sj} m_{ij}+1;-\frac{m_{ij}\Omega_{sj}}{m_{sj} \Omega_{ij} x}\right)$, $\varphi_{21} (x)=(x+1)^{s_2+1} x^{m_{sj}+s_1} \left(\frac{m_{sj} \Omega_{ij}}{ m_{sj} \Omega_{ij} x+m_{ij}\Omega_{sj}}\right)^{I^{sj} m_{ij}+m_{sj}} \\\left(\frac{m_{jd} \Omega_{vj}}{ m_{jd} \Omega_{vj} x+m_{vj}\Omega_{jd}}\right)^{I^{jd} m_{vj}+s_1+s_2+1}\, _2F_1 (I^{sj} m_{ij}+m_{sj},I^{jd} m_{vj}+s_1+s_2+1; f_2+I^{sj} m_{ij}+m_{sj}+I^{jd} m_{vj};1-\frac{m_{jd} m_{sj} \Omega_{vj} \Omega_{ij} x (x+1)}{ \left(\Omega_{jd}m_{vj} m_{ij}+m_{jd} \Omega_{vj} m_{ij} x\right) \left(\Omega_{sj}m_{vj} m_{ij}+\Omega_{ij} m_{sj} xm_{vj}\right)})$.

Again, (\ref{ri31}) is a very good closed-form approximation for the exact CDF when INR is large in the case of i.i.d fixed interferences. Using (\ref{www2}), the outage probability is obtained. Simulations in Fig. \ref{approfix} in Section \uppercase\expandafter{\romannumeral5} will show that this approximation has a very good match with the exact outage probability when INR is large or interferences are dominant.

\subsection{When the signal experiences Rayleigh fading}
In another special case when the signal experiences Rayleigh fading channel, (\ref{ri31}) is further specialized to
\begin{equation}\label{ri32}
\begin{aligned}
&F_{\Gamma_j}(x)=1- I^{sj} m_{ij} \Omega_{jd}^{I^{jd} m_{vj}} \Omega_{sj}^{I^{sj} m_{ij}} B(I^{jd} m_{vj}+1,I^{sj} m_{ij}) \left(\frac{\Omega_{vj} x}{m_{vj}}+\Omega_{jd}\right)^{-I^{jd} m_{vj}} \\& \left(\frac{\Omega_{ij} x}{m_{ij}}+\Omega_{sj}\right)^{-I^{sj} m_{ij}}\varphi_{22} (x) \, _2F_1\left(I^{jd} m_{vj}+1,I^{sj} m_{ij}+1;I^{sj} m_{ij}+I^{jd} m_{vj}+1;1-\varphi_{22} (x)\right)
\end{aligned}
\end{equation}
where
$\varphi_{22} (x)=\frac{\Omega_{ij} \Omega_{vj} x (x+1)}{(m_{ij} \Omega_{sj}+\Omega_{ij} x) (m_{vj} \Omega_{jd}+\Omega_{vj} x)}$.

\noindent  In the high SINR condition such that $\Omega_{sj}\rightarrow \infty$ and $\Omega_{jd}\rightarrow \infty$,
using \cite[(9.1)]{Gradshteyn} as
\begin{equation}\label{ri33}
\begin{aligned}
\lim_{\varphi\to 0} \, \varphi \cdot \, _2F_1(a+1,b+1;a+b+1;1-\varphi)=\frac{\Gamma (a+b+1)}{\Gamma (a+1) \Gamma (b+1)}
\end{aligned}
\end{equation}
in (\ref{ri32}), one can get the high SINR approximation of (\ref{ri32}) as
\begin{equation}\label{ri34}
\begin{aligned}
F_{\Gamma_j}(x)=1-\left(\frac{\Omega_{ij} x}{\Omega_{sj}m_{ij}}+1\right)^{-I^{sj} m_{ij}}\left(\frac{\Omega_{vj} x}{\Omega_{jd} m_{vj}}+1\right)^{-I^{jd} m_{vj}}.
\end{aligned}
\end{equation}
Using (\ref{www2}), the outage probability is obtained. From (\ref{ri34}) and (\ref{www2}), one has several insights as follows:
(1) with the increase of the number of interferers at the relay $I^{sj}$ or at the destination $I^{jd}$, the outage probability increases; (2) with the increase of the average power of signal at the relay $\Omega_{sj}$ or at the destination $\Omega_{jd}$, the outage probability decreases; (3) with the increase of the average power of interferers at the relay $\Omega_{ij}$ or at the destination $\Omega_{vj}$, the outage probability increases.
\section{Numerical Results and Discussion}
\label{six}
In this section, numerical examples are presented to show the effects of the number and locations of interferers by using the outage probability expressions derived in the previous sections.
We assume $\bar{\gamma}^{SINR}=b_{sj}\bar{\gamma}^{SINR}_{sj}=b_{jd}\bar{\gamma}^{SINR}_{jd}$,  $\bar{\gamma}^{INR}=c_{sj}\bar{\gamma}^{INR}_{sj}=c_{jd}\bar{\gamma}^{INR}_{jd}$, $\beta=\beta_{ij}=\beta_{vj}$ and $K_{ij}P_{ij}=K_{vj}P_{vj}=1$.
In the examples where the interferers have random number and locations, we let $\lambda=\lambda_I A_I$ and assume that the distances $l_{ij}$ and $l_{vj}$ follow the uniform distribution as $f_l(l)=\frac{2
l}{L^2}$, $0<l<L$, where $L$ is the maximum radius of the disc. We assume $\lambda=\lambda_{sj}=\lambda_{jd}$ and $L=L_{sj}=L_{jd}$ in this case.
In the examples where the interferers have fixed number and locations, the distances $l_{ij}$, $l_{vj}$ and the number of interferers $I^{sj}$, $I^{jd}$ are constant. Therefore,
we assume $l=l_{ij}=l_{vj}=2$, $I=I^{sj}=I^{jd}=10$ and $\beta=3$. In this case, there still exists path loss if the distances between nodes are large. The path loss is determined by $l$ and $\beta$.
This influence can be checked by examining $\Omega_{ij}$ and $\Omega_{vj}$, as $l$ and $\beta$ can be absorbed by $\Omega_{ij}$ and $\Omega_{vj}$ as part of the average powers of the interference.
In the calculations for both cases above, we assume the number of relays $J=2$, $b_{sj}=c_{sj}=c_{jd}=1$ and $b_{jd}=10$.
Also, let the values of $m_{sj}$ be the same for any $j$; $m_{jd}$ be the same for any $j$; $m_{ij}$ be the same for any $i$, $j$; $m_{vj}$ be the same for any $v$, $j$.
Note that our results are general enough to include other cases but these settings are used here as examples.

\begin{figure}[htbp!]
\centering
\includegraphics[width=3.8in]{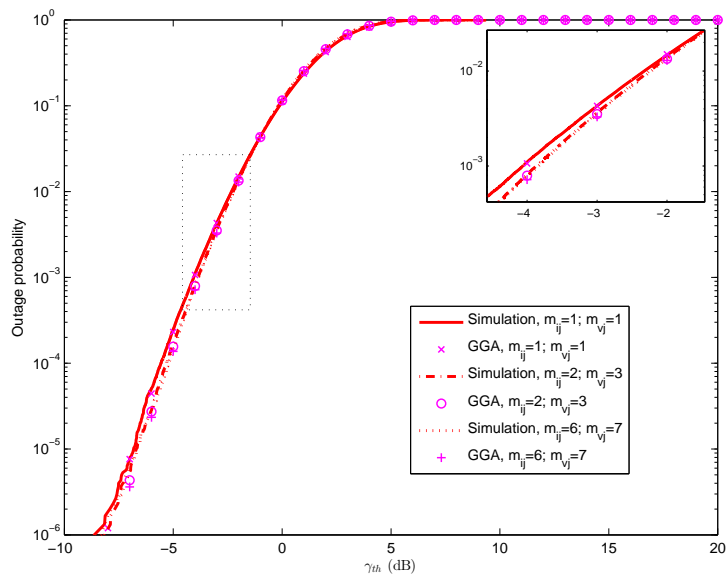}
\caption{\normalsize{Outage probability vs. $\gamma_{th}$ for random interferers when $\bar{\gamma}^{SINR}$ $=15$ dB, $\bar{\gamma}^{INR}$ $=0$ dB, $\lambda=50$, $L=10$, $\beta=3$, $m_{sj}=4$ and $m_{jd}=5$.}}
\label{mij}
\end{figure}

\begin{figure}[htbp!]
\centering
\includegraphics[width=3.8in]{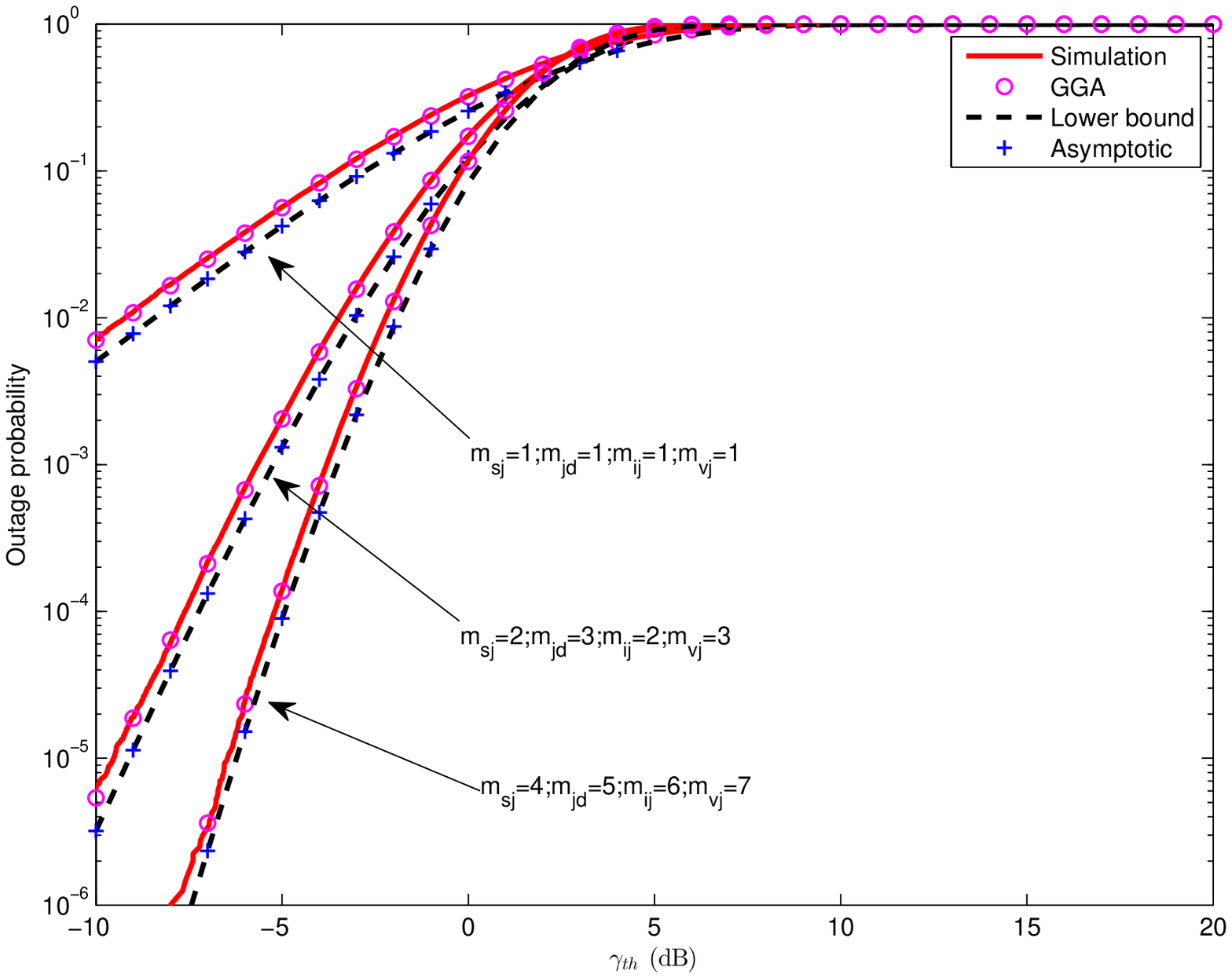}
\caption{\normalsize{Outage probability vs. $\gamma_{th}$ for random interferers when $\bar{\gamma}^{SINR}$ $=15$ dB, $\bar{\gamma}^{INR}$ $=0$ dB, $\lambda=50$, $L=10$ and $\beta=3$.}}
\label{SNR=15nINR=0}
\end{figure}

\begin{figure}[htbp!]
\centering
\includegraphics[width=3.8in]{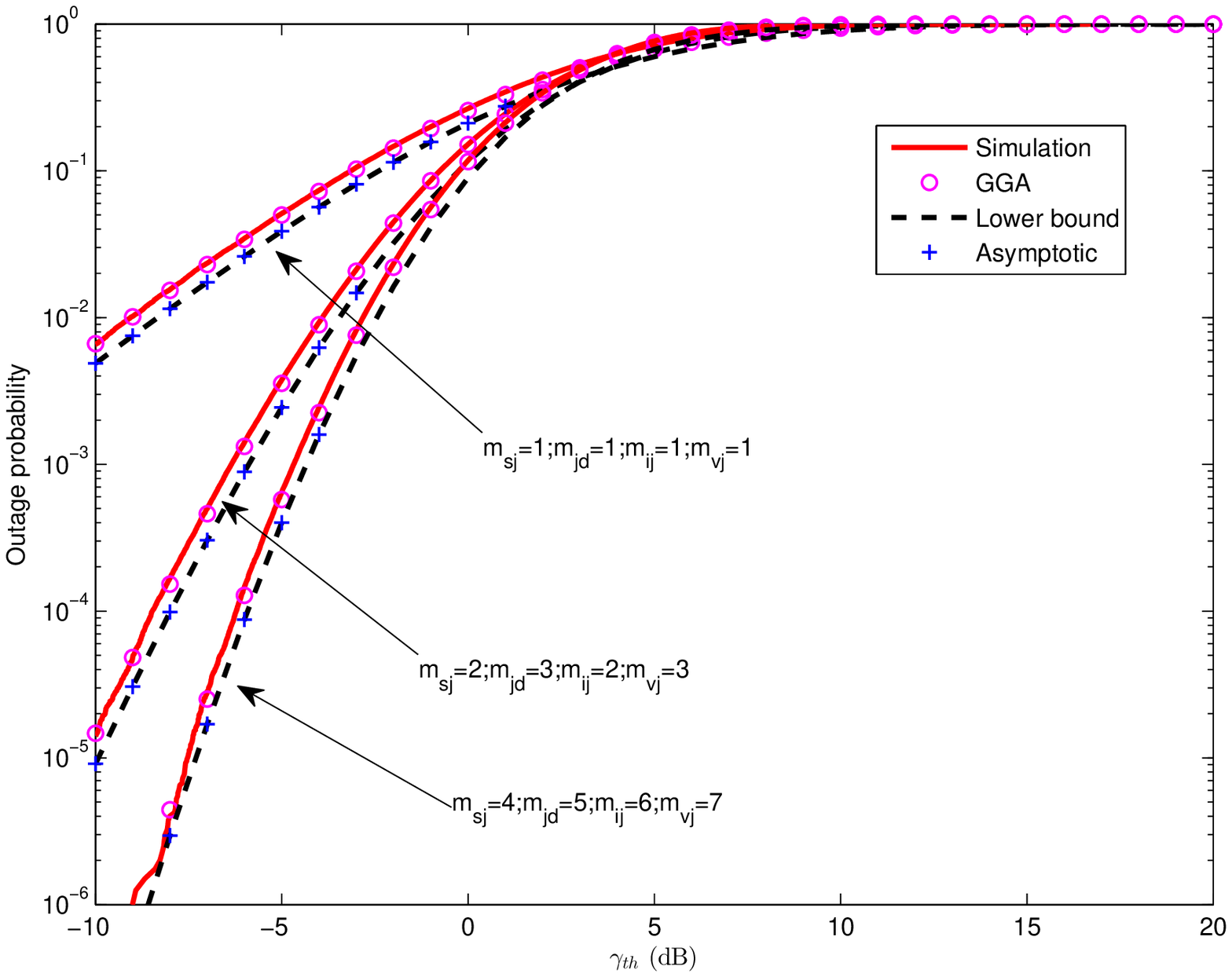}
\caption{\normalsize{Outage probability vs. $\gamma_{th}$ for random interferers when $\bar{\gamma}^{SINR}$ $=15$ dB, $\bar{\gamma}^{INR}$ $=20$ dB, $\lambda=50$, $L=10$ and $\beta=3$.}}
\label{SNR=15nINR=20}
\end{figure}

\begin{figure}[htbp!]
\centering
\includegraphics[width=3.8in]{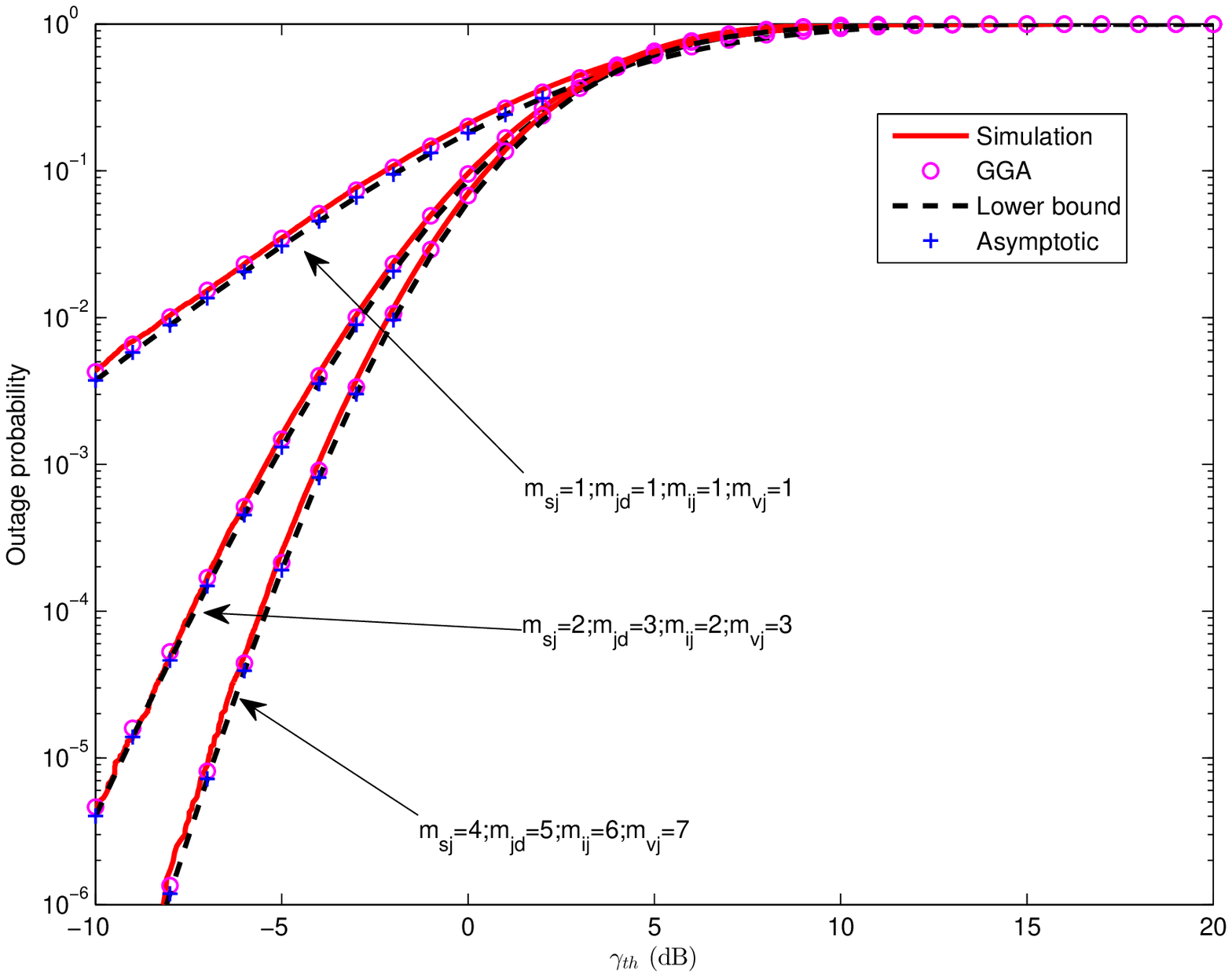}
\caption{\normalsize{Outage probability vs. $\gamma_{th}$ for random interferers when $\bar{\gamma}^{SINR}$ $=20$ dB, $\bar{\gamma}^{INR}$ $=20$ dB, $\lambda=50$, $L=10$ and $\beta=3$.}}
\label{SNR=20nINR=20}
\end{figure}

\begin{figure}[htbp!]
\centering
\includegraphics[width=3.8in]{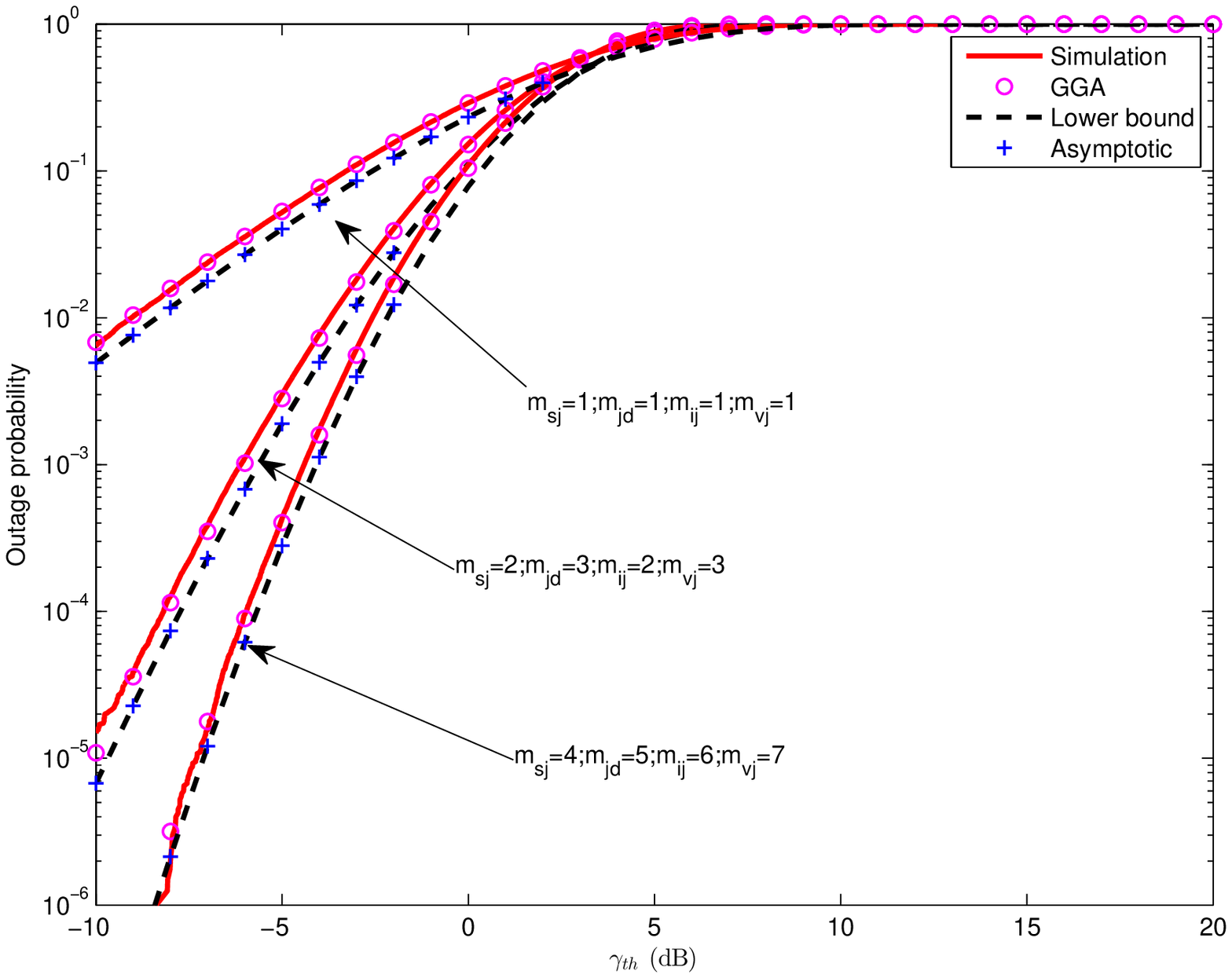}
\caption{\normalsize{Outage probability vs. $\gamma_{th}$ for random interferers when $\bar{\gamma}^{SINR}$ $=15$ dB, $\bar{\gamma}^{INR}$ $=0$ dB, $\lambda=50$, $L=10$ and $\beta=5$.}}
\label{theta5}
\end{figure}

\begin{figure}[htbp!]
\centering
\includegraphics[width=3.8in]{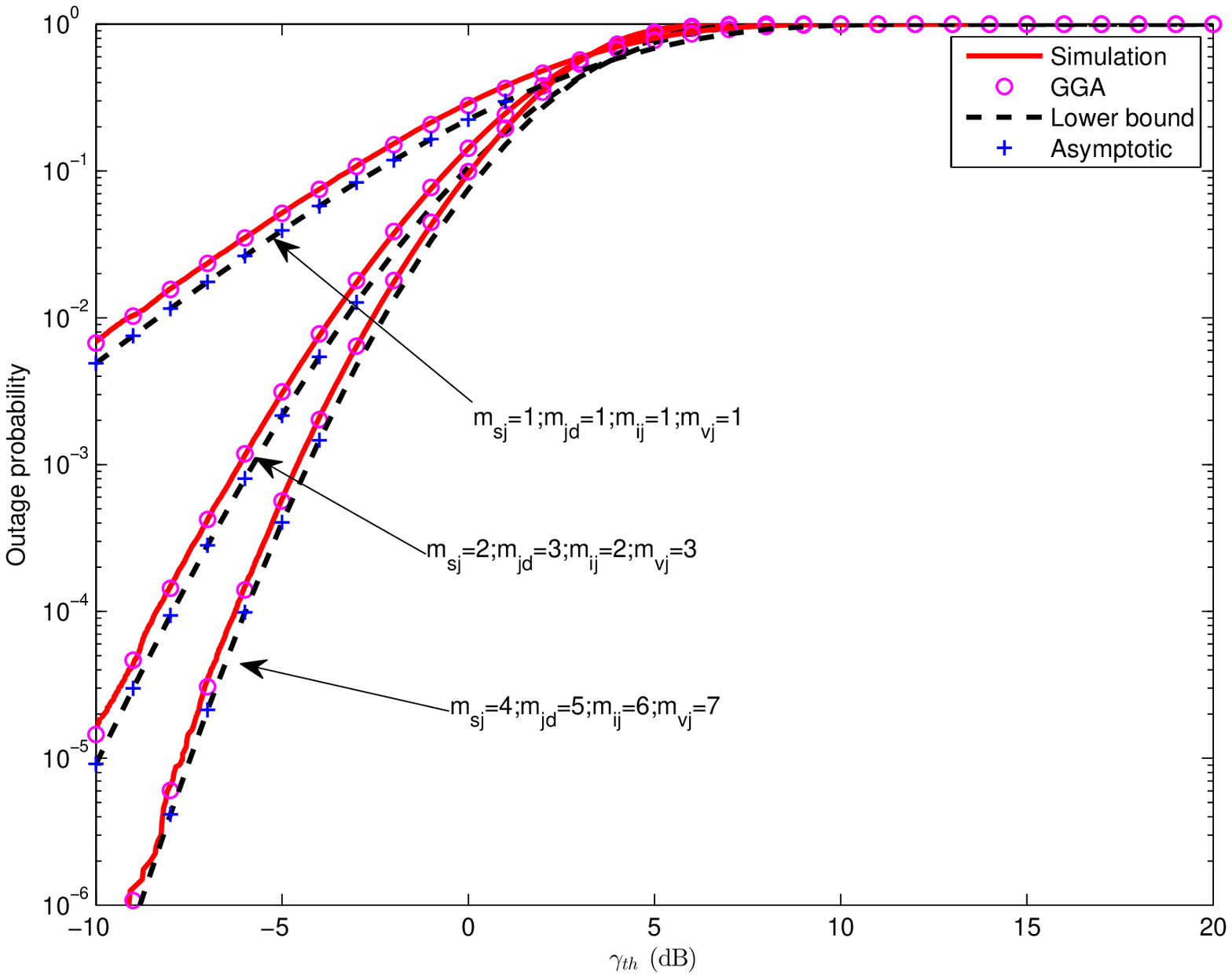}
\caption{\normalsize{Outage probability vs. $\gamma_{th}$ for random interferers when $\bar{\gamma}^{SINR}$ $=15$ dB, $\bar{\gamma}^{INR}$ $=0$ dB, $\lambda=50$, $L=20$ and $\beta=3$.}}
\label{L20}
\end{figure}

Figs.\;\ref{mij}\;-\;\ref{L20} show the outage probability vs.
$\gamma_{th}$ in the case when the interferers have random number and locations.
The GGA curve is obtained by using (\ref{ri15}), (\ref{ri16}) and (\ref{ri19}) in (\ref{www2}) with numerical integration, the lower bound curve is obtained by using (\ref{ri21}) in (\ref{www2})
while the asymptotic curve is obtained by using (\ref{ri22}) in (\ref{www2}).
In general, one can see that the outage probability decreases when the value of $\gamma_{th}$ decreases or when the channel condition changes from Rayleigh fading to general Nakagami-$m$ fading (or with the increase of $m$ in Nakagami-$m$ fading).
The influence of $m_{ij}$ and $m_{vj}$ is examined in
Fig.\;\ref{mij} for $\bar{\gamma}^{SINR}$ $=15$ dB, $\bar{\gamma}^{INR}$ $=0$ dB, $\lambda=50$, $\beta=3$, $L=10$, $m_{sj}=4$ and $m_{jd}=5$. One can see that the curves for $m_{ij}=1$ and $m_{vj}=1$ have a slightly worse outage probability than the curves for $m_{ij}=2$, $m_{vj}=3$ and $m_{ij}=6$, $m_{vj}=7$ while the curves for $m_{ij}=2$, $m_{vj}=3$ and $m_{ij}=6$, $m_{vj}=7$ are nearly the same for the reasons explained below (\ref{ri22}).
Fig.\;\ref{SNR=15nINR=0} shows the
result for $\bar{\gamma}^{SINR}$ $=15$ dB, $\bar{\gamma}^{INR}$ $=0$ dB, $\lambda=50$, $\beta=3$,  and $L=10$ while
Fig.\;\ref{SNR=15nINR=20} shows the result for the same conditions except that
$\bar{\gamma}^{INR}$ is increased from $0$ dB in Fig.\;\ref{SNR=15nINR=0} to $20$ dB in
Fig.\;\ref{SNR=15nINR=20}. One can see the outage probability for $m_{sj}=2$, $m_{jd}=3$, $m_{sj}=2$, $m_{jd}=3$ and $m_{sj}=4$, $m_{jd}=5$, $m_{sj}=6$, $m_{jd}=7$ deteriorate when $\bar{\gamma}^{INR}$ increases and the deteriorate rate for $m_{sj}=2$, $m_{jd}=3$, $m_{sj}=2$, $m_{jd}=3$ is slightly smaller than that for $m_{sj}=4$, $m_{jd}=5$, $m_{sj}=6$, $m_{jd}=7$. However, the outage probability remains nearly unchanged for the Rayleigh case in these two figures for the reasons explained below (\ref{ri22}).
Fig.\;\ref{SNR=20nINR=20} shows the same conditions as Fig.\;\ref{SNR=15nINR=20} except $\bar{\gamma}^{SINR}$ is increased from 15 dB in Fig.\;\ref{SNR=15nINR=20} to 20 dB in Fig.\;\ref{SNR=20nINR=20}. One can see that the outage probability decreases with the increase of $\bar{\gamma}^{SINR}$, as expected. Also,
comparing Fig.\;
\ref{SNR=15nINR=20} with Fig.\;\ref{theta5}, one can see that the outage probability for $m_{sj}=2$, $m_{jd}=3$, $m_{sj}=2$, $m_{jd}=3$ and $m_{sj}=4$, $m_{jd}=5$, $m_{sj}=6$, $m_{jd}=7$ increases when the value of $\beta$ changes from 3 in Fig.\;\ref{SNR=15nINR=20} to 5 in Fig.\;\ref{theta5}.
Comparing Fig.\;
\ref{SNR=15nINR=20} with Fig.\;\ref{L20}, one can see that the outage probability for $m_{sj}=2$, $m_{jd}=3$, $m_{sj}=2$, $m_{jd}=3$ and $m_{sj}=4$, $m_{jd}=5$, $m_{sj}=6$, $m_{jd}=7$ increases when the value of $L$ increases from 10 in Fig.\;\ref{SNR=15nINR=20} to 20 in Fig.\;\ref{L20}.
However, the outage performances for the Rayleigh case in these two cases above keep almost unchanged for the reasons explained below (\ref{ri22}).

In all these cases above, the results based on GGA match very well with the simulation results, showing the accuracy of the approximation and the usefulness of our results.
Also, from Figs.\;\ref{SNR=15nINR=0}\;-\;\ref{L20}, one can see that the lower bounds have considerable match with the simulation while the asymptotic curves match well with the simulation for small $\gamma_{th}$.
On the other hand, one can see that
the gap between lower bound and simulation decreases when $\bar{\gamma}^{SINR}$ increases, as expected, when comparing Fig.\;\ref{SNR=20nINR=20} with other figures above.

\begin{figure}[htbp!]
\centering
\includegraphics[width=3.8in]{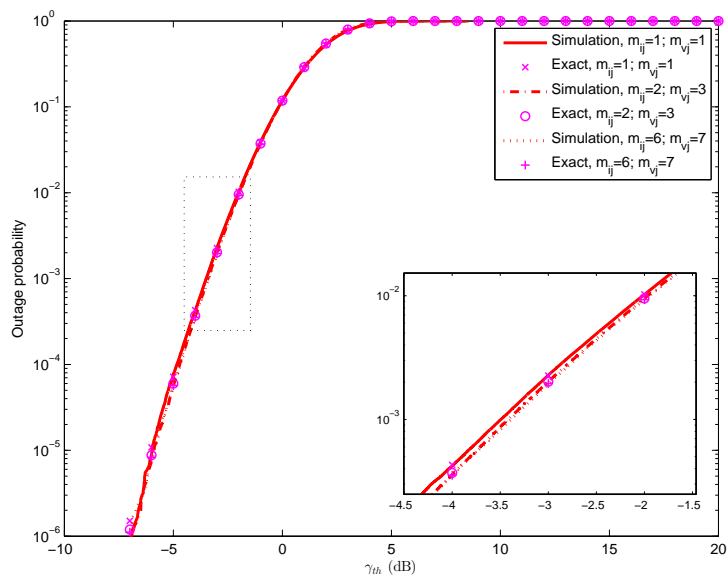}
\caption{\normalsize{Outage probability vs. $\gamma_{th}$ for fixed interferers when $\bar{\gamma}^{SINR}$ $=15$ dB, $\bar{\gamma}^{INR}$ $=0$ dB, $m_{sj}=4$, $m_{jd}=5$.}}
\label{mijmvj}
\end{figure}

\begin{figure}[htbp!]
\centering
\includegraphics[width=3.8in]{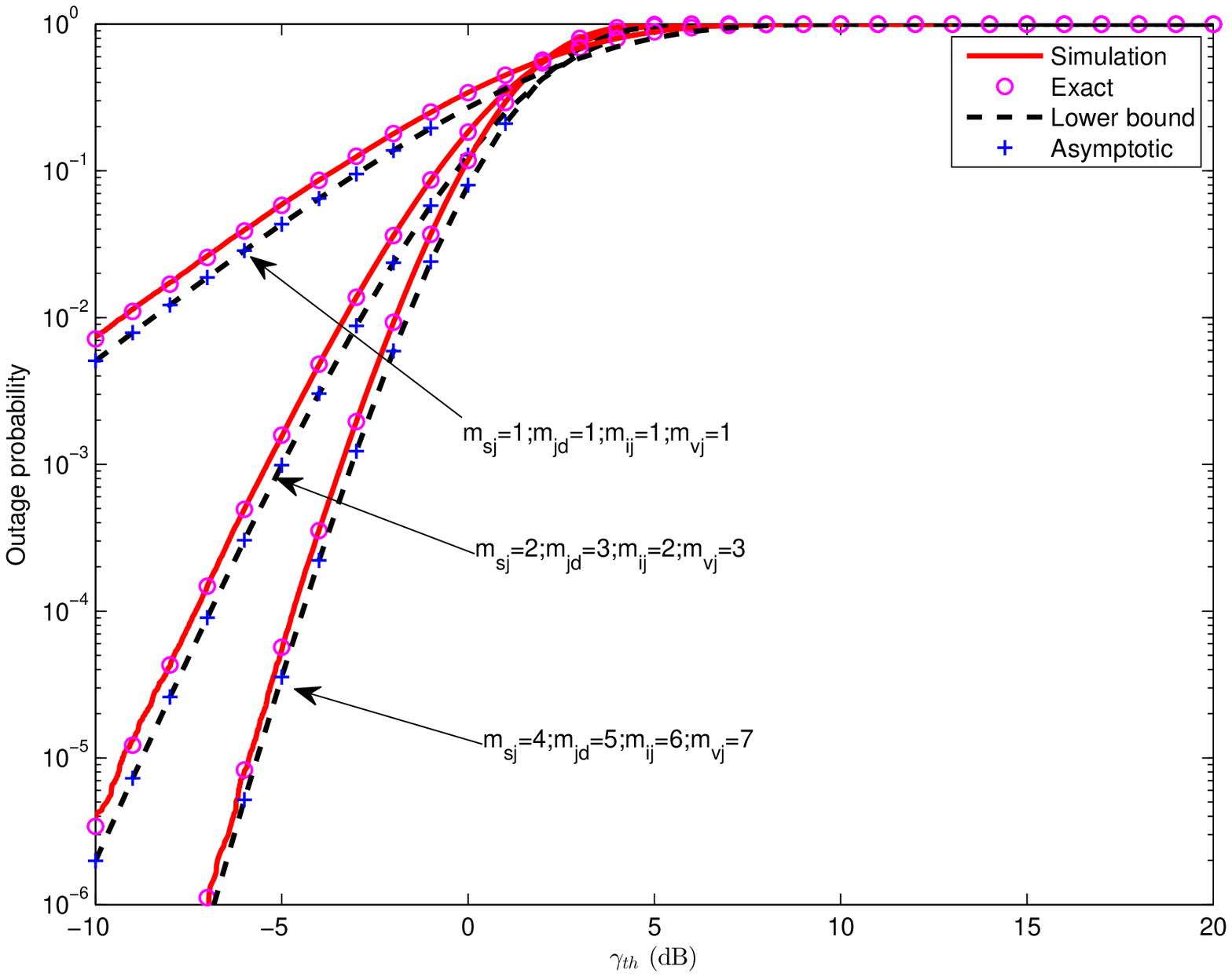}
\caption{\normalsize{Outage probability vs. $\gamma_{th}$ for fixed interferers when $\bar{\gamma}^{SINR}$ $=15$ dB, $\bar{\gamma}^{INR}$ $=0$ dB.}}
\label{s15n0}
\end{figure}

\begin{figure}[htbp!]
\centering
\includegraphics[width=3.8in]{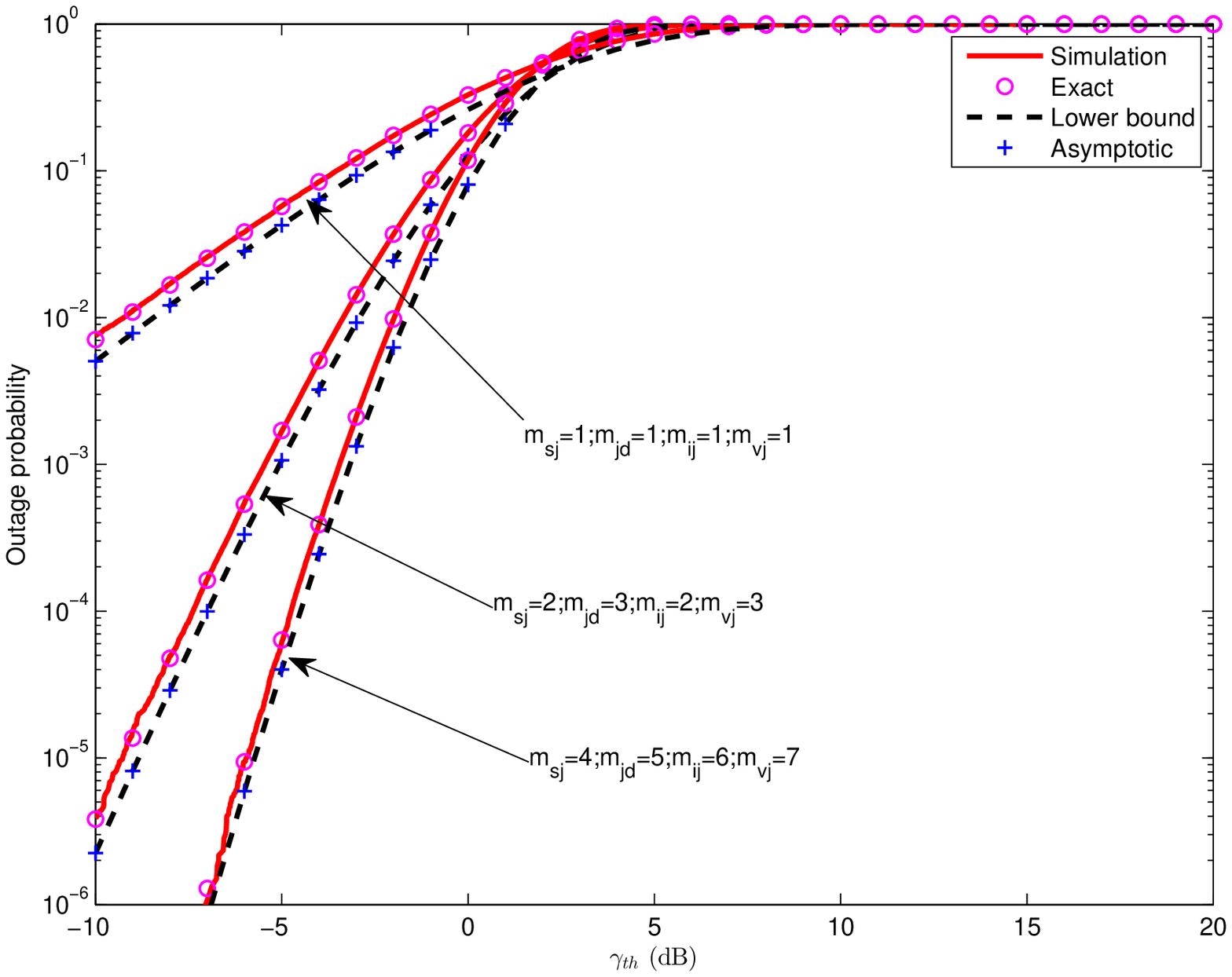}
\caption{\normalsize{Outage probability vs. $\gamma_{th}$ for fixed interferers when $\bar{\gamma}^{SINR}$ $=15$ dB, $\bar{\gamma}^{INR}$ $=20$ dB.}}
\label{s15n20}
\end{figure}

\begin{figure}[htbp!]
\centering
\includegraphics[width=3.8in]{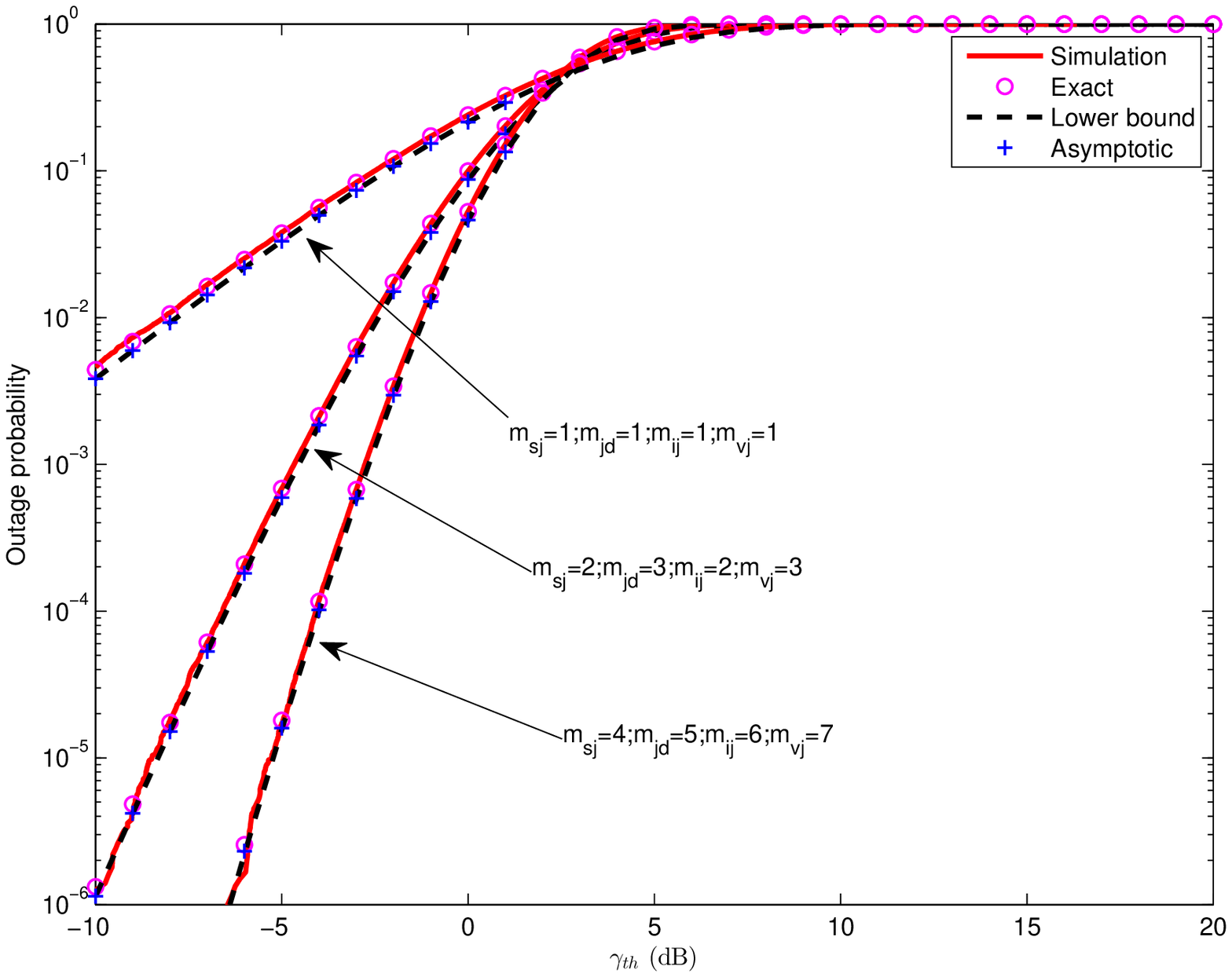}
\caption{\normalsize{Outage probability vs. $\gamma_{th}$ for fixed interferers when $\bar{\gamma}^{SINR}$ $=20$ dB, $\bar{\gamma}^{INR}$ $=20$ dB.}}
\label{s20n20}
\end{figure}

\begin{figure}[htbp!]
\centering
\includegraphics[width=3.8in]{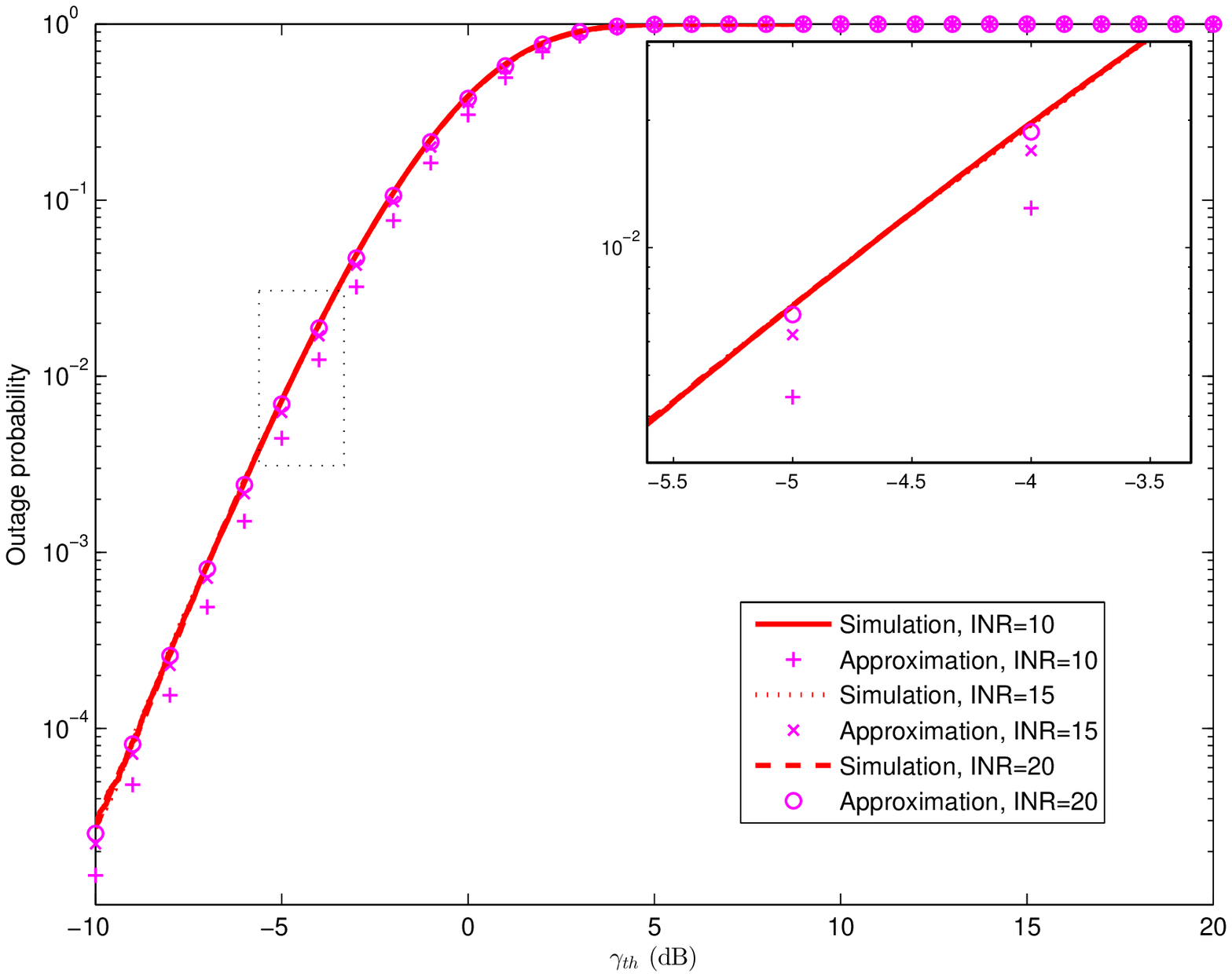}
\caption{\normalsize{Outage probability vs. $\gamma_{th}$ for fixed interferers when $\bar{\gamma}^{SINR}$ $=10$ dB, $m_{sj}=2$, $m_{jd}=3$, $m_{ij}=2$ and $m_{vj}=3$.}}
\label{approfix}
\end{figure}

Figs. \ref{mijmvj} - \ref{approfix} show the outage probability vs. $\gamma_{th}$ in the case when the interferers have fixed number and locations. The exact curve is obtained by using (\ref{ri19}) (\ref{ri25}) and (\ref{ri26}) in (\ref{www2}) with numerical integration, the lower bound curve
is obtained by using (\ref{ew1}) in (\ref{www2})
while the asymptotic curve is obtained by using (\ref{ew1}) and (\ref{ri43}) in (\ref{www2}). In general, one sees that the outage probability decreases when the value of $\gamma_{th}$ decreases or when the Nakagami-$m$ parameter increases. Also, our derived exact results match very well with the simulation results and our derived lower bounds and asymptotic curves have considerable matches with the simulation, especially for small $\gamma_{th}$ in these figures, which verify the accuracy of our analysis.
The influence of $m_{ij}$ and $m_{vj}$ is examined in Fig.\;\ref{mijmvj} for $\bar{\gamma}^{SINR}$ $=15$ dB, $\bar{\gamma}^{INR}$ $=0$ dB, $m_{sj}=4$ and $m_{jd}=5$. One can see that the curves with $m_{ij}=1$ and $m_{vj}=1$ has a slightly worse outage probability than the curves with $m_{ij}=2$, $m_{vj}=3$ and $m_{ij}=6$, $m_{vj}=7$ while the curves with $m_{ij}=2$, $m_{vj}=3$ are almost the same as the curves with $m_{ij}=6$, $m_{vj}=7$.
Fig.\;\ref{s15n0} shows the
result for $\bar{\gamma}^{SINR}$ $=15$ dB, $\bar{\gamma}^{INR}$ $=0$ dB while
Fig.\;\ref{s15n20} shows the result for the same conditions except that
$\bar{\gamma}^{INR}$ is increased from $0$ dB in Fig.\;\ref{s15n0} to $20$ dB in
Fig.\;\ref{s15n20}. One can see that the outage probability from simulation remains almost unchanged, as the SINR dominates the outage probability in the case of fixed interferers and the influence of changing INR can be ignored in this case.
Fig.\;\ref{s20n20} shows the same conditions as Fig.\;\ref{s15n20} except that $\bar{\gamma}^{SINR}$ is increased from 15 dB in Fig.\;\ref{s15n20} to 20 dB in Fig.\;\ref{s20n20}. One can see that the outage probability decreases with the increase of $\bar{\gamma}^{SINR}$, as expected.

Next, our derived closed-form approximations to the exact outage probability in the case of fixed interferences is examined in Fig. \ref{approfix} where (\ref{ri31}) is used as the approximation curve and $\bar{\gamma}^{SINR}=10$ dB, $m_{sj}=2$, $m_{jd}=3$, $m_{sj}=2$, $m_{jd}=3$. One can see that the simulation curves for INR $\bar{\gamma}^{INR}=10$, $15$ and $20$ dB remain almost unchanged, as $\bar{\gamma}^{SINR}$ is fixed in these curves. One can see that the approximation curve with $\bar{\gamma}^{INR}=10$ dB
is closer to the exact curve in Fig.\;\ref{approfix} but still have a slight approximation error.
With the increase of $\bar{\gamma}^{INR}$, these approximation errors decrease. In the case of $\bar{\gamma}^{INR}=20$ dB, this approximation error can be ignored.

\section{Conclusions}
\label{seven}
The outage probability performance of a dual-hop AF selective relaying system with relay selection based on the global instantaneous SINR has been analyzed for different cases of interferer number and locations. Exact analytical expressions in terms of one-dimensional integral for the general cases have been derived. Closed-form expressions for its lower bounds and asymptotic bounds have been obtained. Special cases for dominant interferences, i.i.d Nakagami-$m$ fading interferers
and Rayleigh fading channel have also been studied.
Numerical examples have been presented to show the accuracy of the analysis by examining the effects of interferences and their locations, which are otherwise not possible using previous results. These examples have confirmed that the outage performance improves when the SINR increases and provided useful insights on the effects of different system parameters on the outage performance.

%

\appendices
\section{Derivation of the PDF of $Y_{sj}$ for random interferers}
One has the moment generation function of
(\ref{ri10}) as \cite{Johnson}
\begin{equation}\label{ri35}
\begin{aligned}
E(Y_{sj}^c)=\frac{a_{sj}^c \Gamma \left(\frac{d_{sj}+c}{p_{sj}}\right)}{\Gamma
\left(\frac{d_{sj}}{p_{sj}}\right)}
\end{aligned}
\end{equation}
where $c$ represents the $c$-th order moment.
Denote $\eta_{1,ij}=\int_{0}^{\infty}f_l(l_{ij})\eta (l_{ij})dl_{ij}$, \\ $\eta_{2,ij}=\int_{0}^{\infty}f_l(l_{ij})\eta^2 (l_{ij})dl_{ij}$ and $\eta_{3,ij}=\int_{0}^{\infty}f_l(l_{ij})\eta^3 (l_{ij})dl_{ij}$.
One can derive
the first-order moment of $Y_{sj}$ as
\begin{equation}\label{ri36}
\begin{aligned}
E\{Y_{sj}\}&=\sum _{I=0}^\infty \frac{e^{-\lambda_I A_I} (\lambda_I
A_I)^I }{I!}\sum _{i=1}^I E\{\Omega_{ij}\}E\{\left |
h_{ij}\right|^2\}
=\sum _{I=0}^\infty \frac{e^{-\lambda_I A_I} (\lambda_I A_I)^I
}{I!}\sum _{i=1}^I K_{ij}P_{ij} \eta_{1,ij},
\end{aligned}
\end{equation}
the second-order moment of $Y_{sj}$ as
\begin{equation}\label{ri37}
\begin{aligned}
E\{Y_{sj}^2\}&=\sum _{I=0}^\infty \frac{e^{-\lambda_I A_I} (\lambda_I
A_I)^I }{I!} E\left\{\left(\sum _{i_1=1}^I\Omega_{i_1j} \;\left |
h_{i_1j}\right|^2\right)\left( \sum _{i_2=1}^I\Omega_{i_2j} \;\left
| h_{i_2j}\right|^2 \right)\right\}
\\&=\sum _{I=0}^\infty \frac{e^{-\lambda_I A_I} (\lambda_I A_I)^I
}{I!}\left(\sum _{i=1}^I K^2_{ij}P^2_{ij}\eta_{2,ij}\frac{m_{ij}+1}{m_{ij}}
+
\sum _{i_1=1}^I\sum _{i_2\neq i_1  =1}^I
K_{i_1j}P_{i_1j}K_{i_2j}P_{i_2j}
\eta^2_{1,ij}\right)
\end{aligned}
\end{equation}
and the third-order moment of $Y_{sj}$ as
\begin{equation}\label{ri38}
\begin{aligned}
&E\{Y_{sj}^3\}=\sum _{I=0}^\infty \frac{e^{-\lambda_I A_I} (\lambda_I
A_I)^I }{I!} E\left\{\left(\sum _{i_1=1}^I\Omega_{i_1j} \;\left |
h_{i_1j}\right|^2\right)\left( \sum _{i_2=1}^I\Omega_{i_2j} \;\left
| h_{i_2j}\right|^2 \right)\left(\sum _{i_1=1}^I\Omega_{i_3j}
\;\left | h_{i_3j}\right|^2\right)\right\}
\\&=\sum _{I=0}^\infty \frac{e^{-\lambda_I A_I} (\lambda_I A_I)^I
}{I!}\left(\sum _{i=1}^I
K^3_{ij}P^3_{ij}\eta_{3,ij}\frac{(m_{ij}+1)(m_{ij}+2)}{m_{ij}^2}
+\sum _{i_1=1}^I\sum _{i_2\neq i_1 =1}^I
K^2_{i_1j}P^2_{i_1j}K_{i_2j}P_{i_2j}\right.\\&\left.\eta_{2,ij}\eta_{1,ij} \frac{m_{ij}+1}{m_{ij}}
+ \sum _{i_1=1}^I\sum _{i_2\neq i_1 =1}^I \sum
_{i_3\neq i_2\neq i_1 =1}^I
K_{i_1j}P_{i_1j}K_{i_2j}P_{i_2j}K_{i_3j}P_{i_3j}
\eta^3_{1,ij}\right).
\end{aligned}
\end{equation}
Note that (\ref{ri36}), (\ref{ri37}) and (\ref{ri38}) require an infinite series. However, in reality, one does not need to include many terms in the calculation as $\frac{e^{-\lambda_I A_I} (\lambda_I
A_I)^I }{I!}$ decreases quickly with $I$. Therefore approximations of (\ref{ri36}), (\ref{ri37}) and (\ref{ri38}) can be made by choosing finite series.
Then, one can calculate the values of $a_{sj}$, $p_{sj}$ and $d_{sj}$ in
(\ref{ri10}) by solving
\begin{equation}\label{ri39}
\begin{aligned}
\left\{\begin{matrix}
E\{Y_{sj}\}=\frac{a_{sj} \Gamma \left(\frac{d_{sj}+1}{p_{sj}}\right)}{\Gamma
\left(\frac{d_{sj}}{p_{sj}}\right)}\\E\{Y_{sj}^2\}=\frac{a_{sj}^2 \Gamma \left(\frac{d_{sj}+2}{p_{sj}}\right)}{\Gamma
\left(\frac{d_{sj}}{p_{sj}}\right)}\\E\{Y_{sj}^3\}=\frac{a_{sj}^3 \Gamma \left(\frac{d_{sj}+3}{p_{sj}}\right)}{\Gamma
\left(\frac{d_{sj}}{p_{sj}}\right)}.
\end{matrix}\right.
\end{aligned}
\end{equation}
Furthermore, with the help of Beta
function $B(\cdot, \cdot)$ \cite[(8.384)]{Gradshteyn}, one can simplify (\ref{ri39}) as (\ref{ri11}), that can be solved numerically by using popular mathematical
software packages, such as MATLAB, MATHEMATICA and MAPLE.

\section{Derivation of the PDF of $\Gamma_{sj}$ for random interferers}
Assume independent random variables $u, x>0$ in the equations below.
Using (\ref{ri10}) and (\ref{ri13}) in (\ref{ri14}) and after
some manipulations, one has
\begin{equation}\label{ri40}
\begin{aligned}
f_{\Gamma_{sj}}(u)=\frac{p_{sj}\left(\frac{m_{sj}}{\Omega_{sj}}\right)^{m_{sj}}u^{m_{sj}-1}}{{a_{sj}^{d_{sj}}
\Gamma (m_{sj}) \Gamma
\left(\frac{d_{sj}}{p_{sj}}\right)}}\int_{\sigma_{sj}^2}^{\infty}
(x-\sigma_{sj}^2)^{d_{sj}-1} x^{m_{sj}}  \exp
\left(-\left(\frac{x-\sigma_{sj}^2}{a_{sj}}\right)^{p_{sj}}-\frac{m_{sj}
x u}{\Omega_{sj}}\right)d x.
\end{aligned}
\end{equation}
Using binomial expansion and variable substitution,
(\ref{ri40}) becomes
\begin{equation}\label{ri41}
\begin{aligned}
f_{\Gamma_{sj}}(u)&=\sum _{r_1=0}^{m_{sj}}\frac{p_{sj} m_{sj}!
a_{sj}^{-d_{sj}} \left(\frac{m_{sj}}{\Omega_{sj}
}\right)^{m_{sj}}\sigma_{sj}^{2
(m_{sj}-r_1)}x^{m_{sj}-1}e^{-\frac{m_{sj} \sigma_{sj}^2 u}{\Omega_{sj}
}}}{\Gamma (m_{sj}) \Gamma \left(\frac{d_{sj}}{p_{sj}}\right)r_1!
(m_{sj}-r_1)!} \\&\times\int_{0}^{\infty}
x^{d_{sj}+r_1-1}e^{-\left(\frac{x}{a_{sj}}\right)^{p_{sj}}}
  e^{-\frac{m_{sj} u x}{\Omega_{sj} }}d x.
\end{aligned}
\end{equation}
The integral in (\ref{ri41}) can be transformed by replacing the exponential functions with the Meijer's G-function as \cite[pp. 346]{PrudnikovAP}
\begin{equation}\label{ri42}
\begin{aligned}
f_{\Gamma_{sj}}(u)&=\sum _{r_1=0}^{m_{sj}}\frac{p_{sj} m_{sj}!
a_{sj}^{-d_{sj}} \left(\frac{m_{sj}}{\Omega_{sj}
}\right)^{m_{sj}}\sigma_{sj}^{2(m_{sj}-r_1)}u^{m_{sj}-1}e^{-\frac{m_{sj}
\sigma_{sj}^2 u}{\Omega_{sj} }}}{\Gamma (m_{sj}) \Gamma
\left(\frac{d_{sj}}{p_{sj}}\right)r_1! (m_{sj}-r_1)!}\\&\times
\int_{0}^{\infty} x^{d_{sj}+r_1-1}
G_{0,1}^{1,0}\left(\left(\frac{x}{a_{sj}}\right)^{p_{sj}}|
\begin{array}{c}
 -\\0 \\
\end{array}
\right) G_{0,1}^{1,0}\left(\frac{m_{sj} u x}{\Omega_{sj} }|
\begin{array}{c}
 -\\0 \\
\end{array}
\right)d x.
\end{aligned}
\end{equation}
This integral can be solved by using \cite{Prudnikov} as
(\ref{ri15}).

\section{Derivation of the CDF of $\Gamma_{sj}$ for random interferers}
Using the definition of CDF and (\ref{ri40}), one has

\begin{equation}\label{ri44}
\begin{aligned}
F_{\Gamma_{sj}}(u)&=\int_{0}^{u}f_{\Gamma_1}(t)dt=\frac{p_{sj}\left(\frac{m_{sj}}{\Omega_{sj}}\right)^{m_{sj}}}{{a_{sj}^{d_{sj}}
\Gamma (m_{sj}) \Gamma
\left(\frac{d_{sj}}{p_{sj}}\right)}}\\&\times\int_{0}^{u}\int_{\sigma_{sj}^2}^{\infty}
t^{m_{sj}-1}(x-\sigma_{sj}^2)^{d_{sj}-1} x^{m_{sj}}  \exp
\left(-\left(\frac{x-\sigma_{sj}^2}{a_{sj}}\right)^{p_{sj}}-\frac{m_{sj}
x t}{\Omega_{sj}}\right)dx dt.
\end{aligned}
\end{equation}
By interchanging the order of integration and solving the
integration over $t$ first using \cite[(3.351)]{Gradshteyn}, one can
get
\begin{equation}\label{ri45}
\begin{aligned}
F_{\Gamma_{sj}}(u)&=\frac{p_{sj} x^{d_{sj}-1} }{a_{sj}^{d_{sj}} \Gamma
(m_{sj}) \Gamma \left(\frac{d_{sj}}{p_{sj}}\right)}\int_{0}^{\infty} \exp
\left(-\left(\frac{x}{a_{sj}}\right)^{p_{sj}}\right)\gamma
\left(m_{sj},\frac{m_{sj} (\sigma_{sj}^2  +x)
u}{\Omega_{sj}}\right)d x
\end{aligned}
\end{equation}
where $\gamma(\cdot,\cdot)$ is the lower incomplete Gamma function
\cite{Gradshteyn}. Then, using \cite[(8.352)]{Gradshteyn} to expand
the lower incomplete Gamma function as a finite series, one can get
\begin{equation}\label{ri46}
\begin{aligned}
F_{\Gamma_{sj}}(u)&=\frac{p_{sj} (m_{sj}-1)!
a_{sj}^{-d_{sj}}}{\Gamma (m_{sj}) \Gamma
\left(\frac{d_{sj}}{p_{sj}}\right)} \left(\int_0^{\infty } x^{d_{sj}-1}
e^{-\left(\frac{x}{a_{sj}}\right)^{p_{sj}}} \, dx
- \sum _{r_2=0}^{m_{sj}-1} \sum _{r_3=0}^{r_2}\right.\\&\left.
\frac{\sigma_{sj}^{2(r_2-r_3)} e^{-\frac{m_{sj} \sigma_{sj}^2
u}{\Omega_{sj} }} \left(\frac{m_{sj} u}{\Omega_{sj}
}\right)^{r_2}}{r_3! (r_2-r_3)!} \int_0^{\infty } x^{d_{sj}+r_3-1} e^
{-\left(\frac{x}{a_{sj}}\right)^{p_{sj}}} e^
{-\frac{m_{sj} u x}{\Omega_{sj} }} \, dx\right).
\end{aligned}
\end{equation}
By using \cite[(3.381)]{Gradshteyn} and using the same method as that for
(\ref{ri42}) twice, one can get the CDF of
$\Gamma_{sj}$ as (\ref{ri16}).

\section{Derivation of the high SINR approximations for PDF and CDF of $\Gamma_{sj}$ for random interferers}

Using Taylor's series expansion of (\ref{ri43}) and \cite[(3.381)]{Gradshteyn}
into (\ref{ri41}) and (\ref{ri46}),
one can calculate the PDF and CDF of $\Gamma_{sj}$ as (\ref{ri17}) and (\ref{ri18}), respectively,
\begin{equation}\label{ri17}
\begin{aligned}
&f_{\Gamma_{sj}}(u)= \sum _{n_1=0}^{N_1}\sum _{r_1=0}^{m_{sj}}\mu_{3,sj,r_1,n_1}e^{-\frac{m_{sj} \sigma_{sj}^2 u}{\Omega_{sj}}}u^{m_{sj}+n_1-1}+o\left[(u/\Omega_{sj})^{N_1}\right],
\end{aligned}
\end{equation}
where $\mu_{3,sj,r_1,n_1}=\frac{(-1)^{n_1} m_{sj}^{m_{sj}+n_1+1} \Omega_{sj}^{-m_{sj}-n_1} \sigma_{sj}^{2(m_{sj}-r_1)} a_{sj}^{d_{sj}+n_1+r_1} \Gamma \left(\frac{d_{sj}+n_1+r_1}{p_{sj}}\right)a_{sj}^{-d_{sj}}}{n_1! \Gamma (r_1+1) \Gamma (m_{sj}-r_1+1)\Gamma \left(\frac{d_{sj}}{p_{sj}}\right)}$,
\begin{equation}\label{ri18}
\begin{aligned}
&F_{\Gamma_{sj}}(u)=1-\sum _{n_2=0}^{N_2}\sum _{r_2=0}^{m_{sj}-1}\sum _{r_3=0}^{r_2}\mu_{4,sj,r_2,r_3,n_2}u^{n_2+r_2} e^{-\frac{m_{sj} \sigma_{sj}^2 u}{\Omega_{sj}}}+o\left[(u/\Omega_{sj})^{N_2}\right],
\end{aligned}
\end{equation}
where $\mu_{4,sj,r_2,r_3,n_2}=\frac{(-1)^{n_2} a_{sj}^{n_2+r_3} \sigma_{sj}^{2(r_2-r_3)} \left(\frac{m_{sj}}{\Omega_{sj}}\right)^{n_2+r_2} \Gamma \left(\frac{d_{sj}+n_2+r_3}{p_{sj}}\right)}{n_2! r_3! (r_2-r_3)! \Gamma \left(\frac{d_{sj}}{p_{sj}}\right)}$.

\noindent Furthermore, using (\ref{ri43}) in (\ref{ri17}) and (\ref{ri18}) again, one can get (\ref{wri17}) and (\ref{wri18}).

\section{Derivation of the PDF and CDF of $\Gamma_{sj}$ for fixed interferers}
Similarly, the PDF of $\Gamma_{sj}$ can be calculated using (\ref{ri13}) and (\ref{ri24}) as
\begin{eqnarray}
\label{ri47}
f_{\Gamma_{sj}}(u)&=&\int_{-\infty}^{\infty}\left | x \right
|f_{W_j}\left ( x u \right )f_{Y_{sj}}\left ( x-\sigma_{sj}^2  \right
)dz_j\nonumber\\&=&
\left(\frac{m_{sj}}{\Omega_{sj}}\right)^{m_{sj}}
\left[\prod_{i^*=1}^{I^{sj}}\left(-\frac{\Omega_{i^*j}}{m_{i^*j}}\right)^
{-m_{i^*j}}\right]\sum_{i=1}^{I^{sj}}\sum_{r=1}^{m_{ij}}\frac{(-1)^rb_{ir}
u^{m_{sj}-1}}{\Gamma(m_{sj})(r-1)!}\nonumber \\ &&\int_{\sigma_{sj}^2}^{\infty}z^{m_{sj}}(z-\sigma_{sj}^2)^{r-1}e^{-\frac{m_{sj}}
{\Omega_{sj}}uz-\frac{m_{ij}}{\Omega_{ij}}(z-\sigma_{sj}^2)}dz.
\end{eqnarray}
This integral can be solved by using \cite[(3.351)]{Gradshteyn} as (\ref{ri25}).

Using the definition of CDF and (\ref{ri47}), one has
\begin{eqnarray}
\label{ri48}
F_{\Gamma_{sj}}(u)&=&\int_0^u f_{\Gamma_{sj}}(t)dt=\left(\frac{m_{sj}}{\Omega_{sj}}\right)^{m_{sj}}
\left[\prod_{i^*=1}^{I^{sj}}\left(-\frac{\Omega_{i^*j}}{m_{i^*j}}\right)^{-m_{i^*j}}\right]
\sum_{i=1}^{I^{sj}}\sum_{r=1}^{m_{ij}}\frac{(-1)^rb_{ir}}{\Gamma(m_{sj})(r-1)!}\nonumber \\ &&\int_0^u\int_{\sigma_{sj}^2}^{\infty}t^{m_{sj}-1}z^{m_{sj}}(z-\sigma_{sj}^2)^{r-1}e^{-\frac{m_{sj}}{\Omega_{sj}}tz-\frac{m_{ij}}{\Omega_{ij}}(z-\sigma_{sj}^2)}dzdt.
\end{eqnarray}
By interchanging the order of integration and solving the integration over $t$ first using \cite[(3.351)]{Gradshteyn}, one further has
\begin{eqnarray}
\label{ri49}
F_{\Gamma_{sj}}(u)&=&\left[\prod_{i^*=1}^{I^{sj}}\left(-\frac{\Omega_{i^*j}}{m_{i^*j}}\right)^{-m_{i^*j}}\right]\sum_{i=1}^{I^{sj}}\sum_{r=1}^{m_{ij}}\frac{(-1)^rb_{ir}}{\Gamma(m_{sj})(r-1)!}\nonumber \\ &&\cdot\int_0^{\infty}z'^{r-1}e^{-\frac{m_{ij}}{\Omega_{ij}}z'}\gamma(m_{sj},\frac{m_{sj}}{\Omega_{sj}}(z'+\sigma_{sj}^2)u)dz'.
\end{eqnarray}
Then, one can derive the CDF of $\Gamma_{sj}$ by using \cite[(8.352)]{Gradshteyn} and \cite[(3.381)]{Gradshteyn} in closed-form as (\ref{ri26}).

\section{Derivation of the CDF of $\Gamma_j$ when the interferences are dominant}
When the interference is dominant such that the noise can be
ignored, one further has $\sigma_{sj}^2\approx 0$ in (\ref{ri25}) to give the PDF as
\begin{eqnarray}
\label{ri50}
f_{\Gamma_{sj}}(u)&=&
\sum_{i=1}^{I^{sj}}\sum_{r=1}^{m_{ij}}\varphi_{5,sj,ij,i,r}
\frac{u^{m_{sj}-1}}{(\frac{m_{sj}}{\Omega_{sj}}u+\frac{m_{ij}}
{\Omega_{ij}})^{m_{sj}+r}},
\end{eqnarray}
where $\varphi_{5,sj,ij,i,r}=
\left[\prod_{i^*=1}^{I^{sj}}\left(-\frac{\Omega_{i^*j}}{m_{i^*j}}\right)^{-m_{i^*j}}\right]
\left(\frac{m_{sj}}{\Omega_{sj}}\right)^{m_{sj}}
\frac{(-1)^r b_{ir}\Gamma(m_{sj}+r)}{\Gamma(m_{sj})(r-1)!}$.

\noindent  Also, the CDF of $\Gamma_{sj}$ in (\ref{ri26}) becomes
\begin{eqnarray}
\label{ri51}
F_{\Gamma_{sj}}(u)&=&1-\sum_{i=1}^{I^{sj}}\sum_{r=1}^{m_{ij}}
\sum_{f=0}^{m_{sj}-1}\varphi_{6,sj,ij,i,r,f}\frac{u^f}
{(\frac{m_{sj}}{\Omega_{sj}}u+\frac{m_{ij}}{\Omega_{ij}})^{f+r}},
\end{eqnarray}
where $\varphi_{6,sj,ij,i,r,f}=\left[\prod_{i^*=1}^{I^{sj}}\left(-\frac{\Omega_{i^*j}}{m_{i^*j}}
\right)^{-m_{i^*j}}\right]\left(\frac{m_{sj}}{\Omega_{sj}}\right)^{f}\frac{(-1)^rb_{ir}
\Gamma(f+r)}{(r-1)!f!}$.
Therefore, (\ref{ri19}) can be solved as (\ref{ri27}) with the help of \cite[(3.197)]{Gradshteyn}.

\bibliographystyle{ieeetran}
\bibliography{relayselection_sg}

\begin{thebibliography}{10}
\providecommand{\url}[1]{#1}
\csname url@samestyle\endcsname
\providecommand{\newblock}{\relax}
\providecommand{\bibinfo}[2]{#2}
\providecommand{\BIBentrySTDinterwordspacing}{\spaceskip=0pt\relax}
\providecommand{\BIBentryALTinterwordstretchfactor}{4}
\providecommand{\BIBentryALTinterwordspacing}{\spaceskip=\fontdimen2\font plus
\BIBentryALTinterwordstretchfactor\fontdimen3\font minus
  \fontdimen4\font\relax}
\providecommand{\BIBforeignlanguage}[2]{{%
\expandafter\ifx\csname l@#1\endcsname\relax
\typeout{** WARNING: IEEEtran.bst: No hyphenation pattern has been}%
\typeout{** loaded for the language `#1'. Using the pattern for}%
\typeout{** the default language instead.}%
\else
\language=\csname l@#1\endcsname
\fi
#2}}
\providecommand{\BIBdecl}{\relax}
\BIBdecl

\bibitem{904590}
J.~Laneman and G.~Wornell, ``Energy-efficient antenna sharing and relaying for
  wireless networks,'' in \emph{2000 IEEE Wireless Communications and
  Networking Confernce}, vol.~1, 2000, pp. 7--12.

\bibitem{1374901}
M.~Hasna and M.-S. Alouini, ``A performance study of dual-hop transmissions
  with fixed gain relays,'' \emph{IEEE Transactions on Wireless
  Communications}, vol.~3, no.~6, pp. 1963--1968, Nov 2004.

\bibitem{5876342}
Y.~Chen, C.-X. Wang, H.~Xiao, and D.~Yuan, ``Novel partial selection schemes
  for af relaying in {N}akagami- $m$ fading channels,'' \emph{IEEE Transactions
  on Vehicular Technology}, vol.~60, no.~7, pp. 3497--3503, Sept 2011.

\bibitem{4801494}
Y.~Jing and H.~Jafarkhani, ``Single and multiple relay selection schemes and
  their achievable diversity orders,'' \emph{IEEE Transactions on Wireless
  Communications}, vol.~8, no.~3, pp. 1414--1423, March 2009.

\bibitem{4290052}
Y.~Zhao, R.~Adve, and T.~J. Lim, ``Improving amplify-and-forward relay
  networks: optimal power allocation versus selection,'' \emph{IEEE
  Transactions on Wireless Communications}, vol.~6, no.~8, pp. 3114--3123,
  August 2007.

\bibitem{4020532}
------, ``Symbol error rate of selection amplify-and-forward relay systems,''
  \emph{IEEE Communications Letters}, vol.~10, no.~11, pp. 757--759, November
  2006.

\bibitem{1603719}
A.~Bletsas, A.~Khisti, D.~Reed, and A.~Lippman, ``A simple cooperative
  diversity method based on network path selection,'' \emph{IEEE Journal on
  Selected Areas in Communications}, vol.~24, no.~3, pp. 659--672, March 2006.

\bibitem{4489652}
I.~Krikidis, J.~Thompson, S.~McLaughlin, and N.~Goertz, ``Amplify-and-forward
  with partial relay selection,'' \emph{IEEE Communications Letters}, vol.~12,
  no.~4, pp. 235--237, April 2008.

\bibitem{6555163}
A.~Guidotti, V.~Buccigrossi, M.~Di~Renzo, G.~Corazza, and F.~Santucci, ``Outage
  and symbol error probabilities of dual-hop af relaying in a poisson field of
  interferers,'' in \emph{2013 IEEE Wireless Communications and Networking
  Conference}, April 2013, pp. 3704--3709.

\bibitem{6130607}
H.~Suraweera, D.~Michalopoulos, and C.~Yuen, ``Performance analysis of fixed
  gain relay systems with a single interferer in {N}akagami-$m$ fading
  channels,'' \emph{IEEE Transactions on Vehicular Technology}, vol.~61, no.~3,
  pp. 1457--1463, March 2012.

\bibitem{6516171}
M.~Di~Renzo, A.~Guidotti, and G.~Corazza, ``Average rate of downlink
  heterogeneous cellular networks over generalized fading channels: A
  stochastic geometry approach,'' \emph{IEEE Transactions on Communications},
  vol.~61, no.~7, pp. 3050--3071, July 2013.

\bibitem{6096455}
Y.~Chen, G.~Karagiannidis, H.~Lu, and N.~Cao, ``Novel approximations to the
  statistics of products of independent random variables and their applications
  in wireless communications,'' \emph{IEEE Transactions on Vehicular
  Technology}, vol.~61, no.~2, pp. 443--454, 2012.

\bibitem{Unnikrishna}
A.~Papoulis and S.~U. Pillai, \emph{Probability, Random Variables and
  Stochastic Processes, 4th ed.}\hskip 1em plus 0.5em minus 0.4em\relax
  McGraw-Hill, 2002.

\bibitem{Gradshteyn}
I.~S. Gradshteyn and I.~M. Ryzhik, \emph{Table of Integrals, Series, and
  Products, 7th ed}.\hskip 1em plus 0.5em minus 0.4em\relax San Diego, CA:
  Academic, 2007.

\bibitem{5426509}
C.~Zhong, S.~Jin, and K.-K. Wong, ``Dual-hop systems with noisy relay and
  interference-limited destination,'' \emph{IEEE Transactions on
  Communications}, vol.~58, no.~3, pp. 764--768, March 2010.

\bibitem{5545638}
H.~Suraweera, H.~Garg, and A.~Nallanathan, ``Performance analysis of two hop
  amplify-and-forward systems with interference at the relay,'' \emph{IEEE
  Communications Letters}, vol.~14, no.~8, pp. 692--694, August 2010.

\bibitem{5740505}
D.~Benevides~da Costa, H.~Ding, and J.~Ge, ``Interference-limited relaying
  transmissions in dual-hop cooperative networks over nakagami-m fading,''
  \emph{IEEE Communications Letters}, vol.~15, no.~5, pp. 503--505, May 2011.

\bibitem{6324367}
F.~Al-Qahtani, J.~Yang, R.~Radaydeh, C.~Zhong, and H.~Alnuweiri, ``Exact outage
  analysis of dual-hop fixed-gain af relaying with cci under dissimilar
  nakagami-m fading,'' \emph{IEEE Communications Letters}, vol.~16, no.~11, pp.
  1756--1759, November 2012.

\bibitem{6189013}
Z.~Lin, X.~Peng, F.~Chin, and W.~Feng, ``Outage performance of relaying with
  directional antennas in the presence of co-channel interferences at relays,''
  \emph{IEEE Wireless Communications Letters}, vol.~1, no.~4, pp. 288--291,
  August 2012.

\bibitem{6209368}
S.~Ikki and S.~Aissa, ``Performance evaluation and optimization of dual-hop
  communication over nakagami-m fading channels in the presence of co-channel
  interferences,'' \emph{IEEE Communications Letters}, vol.~16, no.~8, pp.
  1149--1152, August 2012.

\bibitem{5992706}
D.~da~Costa and M.~Yacoub, ``Outage performance of two hop af relaying systems
  with co-channel interferers over {N}akagami-$m$ fading,'' \emph{IEEE
  Communications Letters}, vol.~15, no.~9, pp. 980--982, September 2011.

\bibitem{5611617}
D.~Lee and J.~H. Lee, ``Outage probability for dual-hop relaying systems with
  multiple interferers over rayleigh fading channels,'' \emph{IEEE Transactions
  on Vehicular Technology}, vol.~60, no.~1, pp. 333--338, Jan 2011.

\bibitem{5784302}
F.~Al-Qahtani, T.~Duong, C.~Zhong, K.~Qaraqe, and H.~Alnuweiri, ``Performance
  analysis of dual-hop af systems with interference in {N}akagami- $m$ fading
  channels,'' \emph{IEEE Signal Processing Letters}, vol.~18, no.~8, pp.
  454--457, Aug 2011.

\bibitem{Mosch}
P.~G. Moschopoulos, ``The distribution of the sum of independent gamma random
  variables,'' \emph{Ann. Inst. Statistical Math. (A)}, vol.~37, pp. 541--544,
  1985.

\bibitem{Mathai}
A.~M. Mathai, ``Storage capacity of a dam with gamma type inputs,'' \emph{Ann.
  Inst. Statist. Math. (A)}, vol.~34, pp. 591--597, 1982.

\bibitem{Johnson}
S.~K. N.~L.~Johnson and N.~Balakrishnan, \emph{Continuous Univariate
  Distributions, 2nd Ed.}\hskip 1em plus 0.5em minus 0.4em\relax Wiley, 1994.

\bibitem{PrudnikovAP}
Y.~A.~B. A.~P.~Prudnikov and O.~I. Marichev, \emph{Integrals and Series, Vol.3:
  More Special Functions}.\hskip 1em plus 0.5em minus 0.4em\relax Oxford
  University Press US, 1986.

\bibitem{Prudnikov}
A.~P. Prudnikov and O.~I. Marichev, ``The algorithm for calculating integrals
  of hypergeometric type functions and its realization in reduce system,'' in
  \emph{Proceedings of the international symposium on Symbolic and algebraic
  computation}, 1990, pp. 212--224.

\end{thebibliography}

\end{document}